\documentclass[a4paper,11pt]{article}
\pdfoutput=1 

\usepackage{jheppub} 

\usepackage[T1]{fontenc} 

\title{\boldmath NLO QCD corrections to the processes pp $\rightarrow $ ZZ with 0, 1 and 2 jets at the LHC.}

\author[a]{K.Djamaa,}
\author[a,b]{A. Mohamed- Meziani}

\affiliation[a]{Department of Physics, Faculty of Exact Sciences, University of Bejaia,06000 Bejaia, Algeria}
\affiliation[b]{Laboratoire de Physique Th\'{e}orique d'Oran (LPTO), University of Oran1-Ahmed Ben Bella; BP 1524 El M'nouar Oran 31000, Algeria}

\emailAdd{kenza.djamaa@univ-bejaia.dz}
\emailAdd{abdelkader.mohamedmeziani@univ-bejaia.dz}

\abstract{  We propose an implementation of ZZ, ZZj and ZZjj productions in MadGraph5\\$\_$aMC@NLO framework at $\sqrt{s}= 14$ TeV. We calculate these processes at leading order and next-to-leading order with QCD corrections and we present a theoretical prediction of their total cross sections with different cuts in transverse momentum of jets, including gluon fusion contributions. In the same time we estimate their theoretical uncertainty. We discuss the various  kinematical distributions spectrums at partonic level and hadronic level applying the showering and hadronization using Pythia8. In order to reconstruct the events similar to that found at LHC, we use the ATLAS cards  and the fast detector simulation Delphes.
	
 }

\begin{document} 
\maketitle
\flushbottom

\section{Introduction }
\label{sec:1}

The production  of vector-boson pairs in proton-proton collisions is one of the most important processes studied at the Large Hadron Collider (LHC), it serves to describe various physics aspects. In particular the Z-boson pairs production plays a crucial role in the test of non-Abelian  gauge structure of electroweak (EW) sector and the mechanism of electroweak symmetry breaking(EWSB). Such process can be used to determine several proprieties of Standard Model (SM) particles at high precision. It was also contribute in the discovery of the Higgs boson, as it is an irreducible background for searching new signals of physics beyond the Standard Model (BSM).

In the last few years, various measurements of ZZ production in proton-proton collisions have been carried out at the ATLAS and CMS Collaborations at $\sqrt{s}= $7 TeV ~\cite{ref:1, ref:2,ref:3}, $\sqrt{s}= $ 8 TeV~\cite{ref:4,ref:5,ref:6} and $\sqrt{s}= $ 13 TeV ~\cite{ref:7,ref:8,ref:9,ref:10}. While measurements of the ZZ production associated to two jets have been studied in both ATLAS and CMS experiments~\cite{ref:11,ref:12}  for $\sqrt{s}= 13$ TeV.

Many theoretical predictions for ZZ production at next-to-leading order (NLO) have been improved  for a long time ago, with both the QCD~\cite{ref:13} and the QED corrections~\cite{ref:14}. The on-shell Z bosons pair production have been studied in~\cite{ref:15,ref:16} from $\sqrt{s}= $ 7 to 14 TeV in the centre of mass, while their leptonic decays have been computed in~\cite{ref:17,ref:13} using Matrix framework and NNPDF3.0 as parton distribution functions with $\alpha_{s}(m_{Z})$ = 0.118. The one-loop induced to gg $\rightarrow $ ZZ calculations has been included in several publications~\cite{ref:18,ref:19,ref:20} matched to parton shower with Pythia8, in addition, the ZZ production via gg fusion mediated by the Higgs boson has been evaluated in~\cite{ref:21,ref:22}. The size of the next-next-to-leading order (NNLO) QCD corrections to this process has been presented in~\cite{ref:17,ref:23}.
In the other hand, the ZZ production associated to jet has his place in the literature, where the on-shell ZZ+ jet production at NLO QCD corrections has been discussed in~\cite{ref:24,ref:25} employing the GOLEM approach, when the off-shell production has been computed in~\cite{ref:26,ref:27}.~\cite{ref:28} gives the next-to-leading order QCD calculations for on-shell ZZ+ jet production via gg fusion. However, the corresponding leptonic decays to the ZZ+ jj in standard model has been predicted in~\cite{ref:29,ref:30,ref:31}. 

 In this paper we report a combination of three processes of pairs bosons creation ZZ with 0, 1 and 2 jets in final state in pp collision. We employ in our simulation MadGraph5$\_$aMC@NLO~\cite{ref:32} to generate the processes at leading-order (LO) and next-to-leading-order (NLO) with QCD corrections. Furthermore, we present the theoretical predictions of the corresponding total cross sections at $\sqrt{s}= $ 14 TeV accompanied with scale uncertainties, for various cuts in transverse momentum of jets ($p_{T}(j) > $ 10, 100 and 200 GeV) in order to understand the effect of these cuts on this magnitude. From these predictions, we can see also the size of the NLO QCD corrections. In addition, we complete our implementation by including the calculation of gluon fusion, which contributes at NLO. 
Our calculation is performed at two levels, first, at partonic level, than at hadronic level matched to parton shower within the Pythia8~\cite{ref:33} framework. Finally  we analyse several distributions ($p_{T}(Z_{1})$, $p_{T}(Z_{2})$, M$_{ZZ}$ and $y[Z_{1} Z_{2}]$ ) using MadAnalysis5~\cite{ref:34}.  

The remainder of the paper is organized as follows. In the next section, we describe the computational details of on-shell Z pair production associated to 0, 1 and 2 jets, at LO and  NLO QCD corrections calculations. Then, in section~\ref{sec:3}, we give the collected input parameters used for our computation, than we present our prediction of the total cross sections and we analyse  various distributions. Finally, we summarise briefly our paper in Section~\ref{sec:4}.

\section{ Details of the calculation}
\label{sec:2}

We consider in this study the production, on shell, of three independents processes:
\begin{eqnarray}
\label{eq:1}
 p p \rightarrow ZZ, \qquad p p \rightarrow ZZj,  \qquad p p \rightarrow ZZjj  
\end{eqnarray}
 in proton-proton collisions at LO and NLO with QCD corrections in the context of the Standard Model. In both levels, the processes are generated with LHC settings with help of MadGraph5$\_$aMC@NLO version 2.6.0. We have implemented our calculations in the 4-flavor scheme, i.e. p = u, d, c, s and gluons and we neglect their masses. We have excluded the bottom, top quarks and Higgs contributions in the initial state but we take in account the virtual bottom-loop, top-loop and Higgs-mediated contributions.

\subsection{Contributions at leading order}    
\label{subsec:1}

\begin{figure}[tbp]
	\centering	
	\includegraphics[width=.30\textwidth,trim=0 0 0 0,clip]{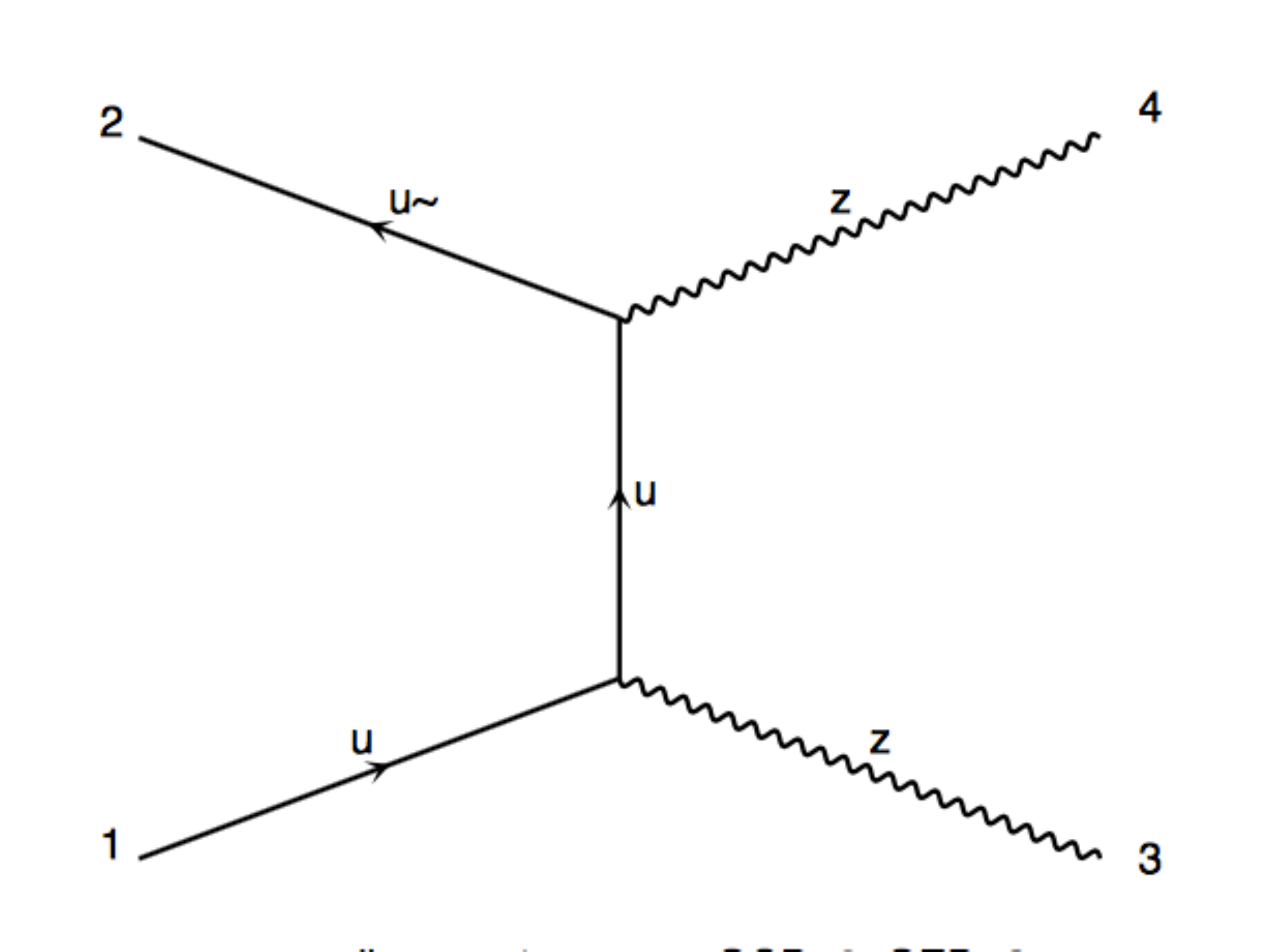}	\includegraphics[width=.30\textwidth,trim=0 0 0 0,clip]{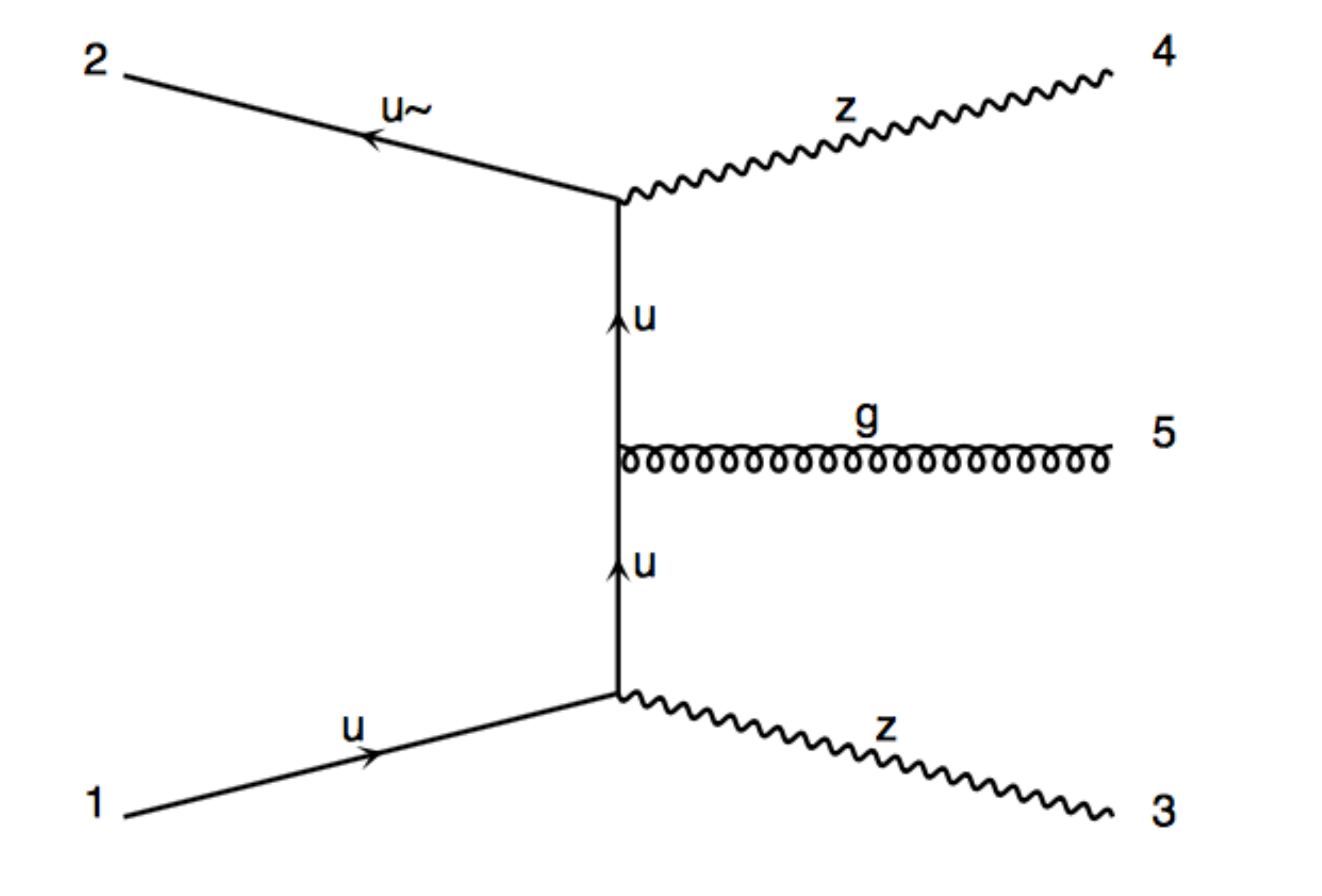} 	
	\includegraphics[width=.30\textwidth,trim=0 0 0 0,clip]{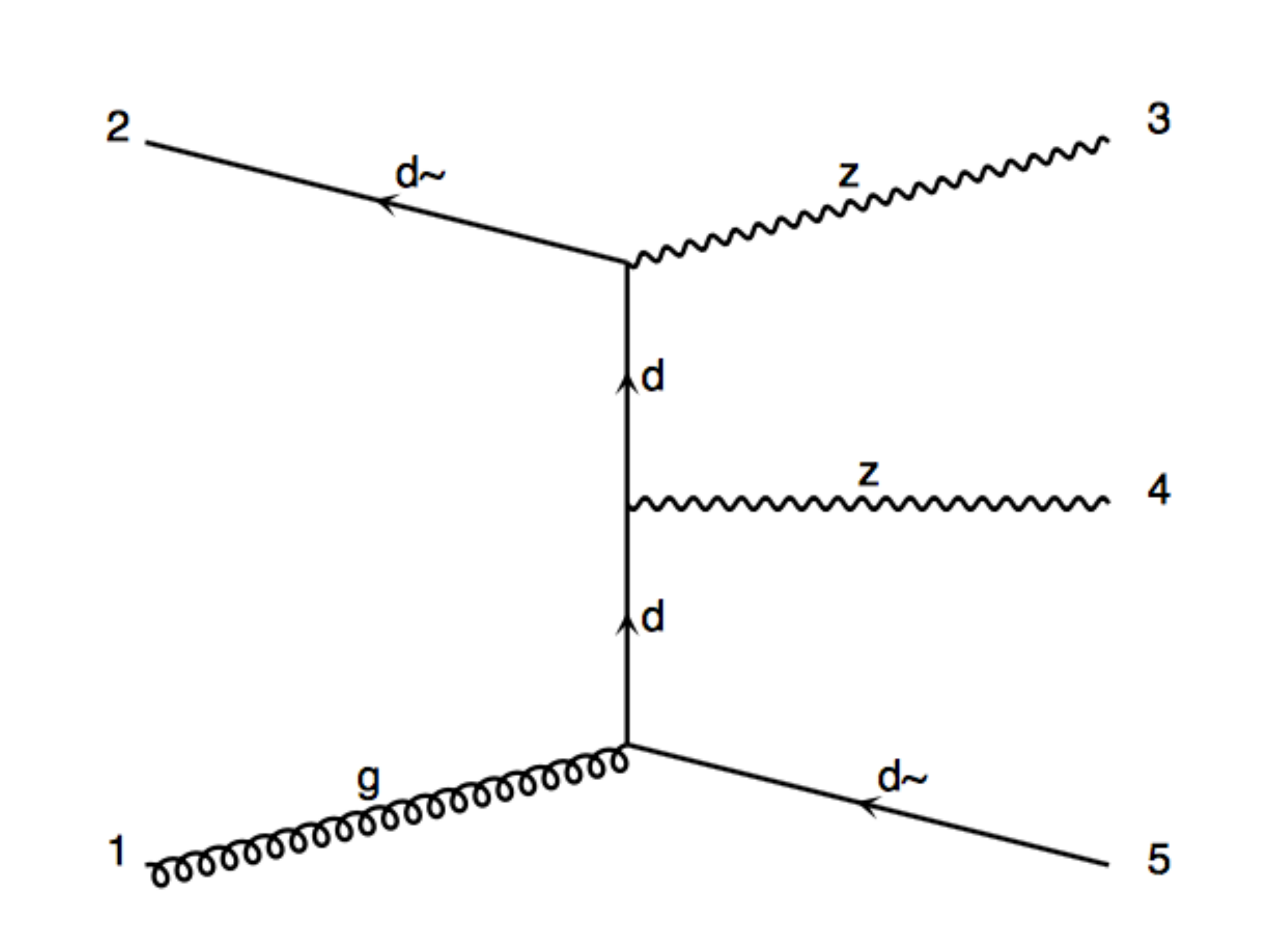} \\
	 a \qquad \qquad \qquad \qquad  \qquad \qquad b  \qquad \qquad \qquad \qquad  \qquad \qquad c \qquad \qquad \qquad \qquad
		\includegraphics[width=.30\textwidth,trim=0 0 0 0,clip]{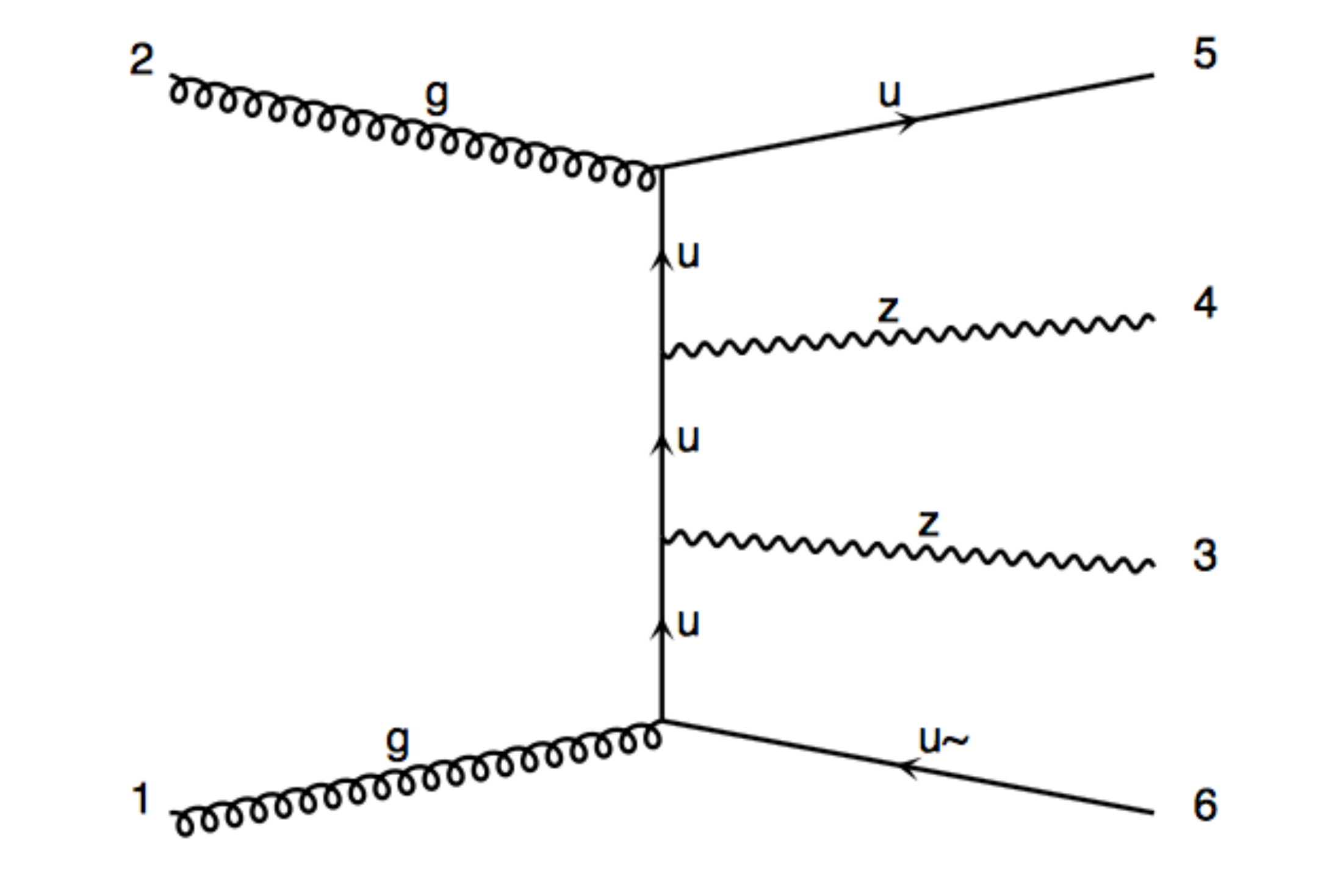}	\includegraphics[width=.30\textwidth,trim=0 0 0 0,clip]{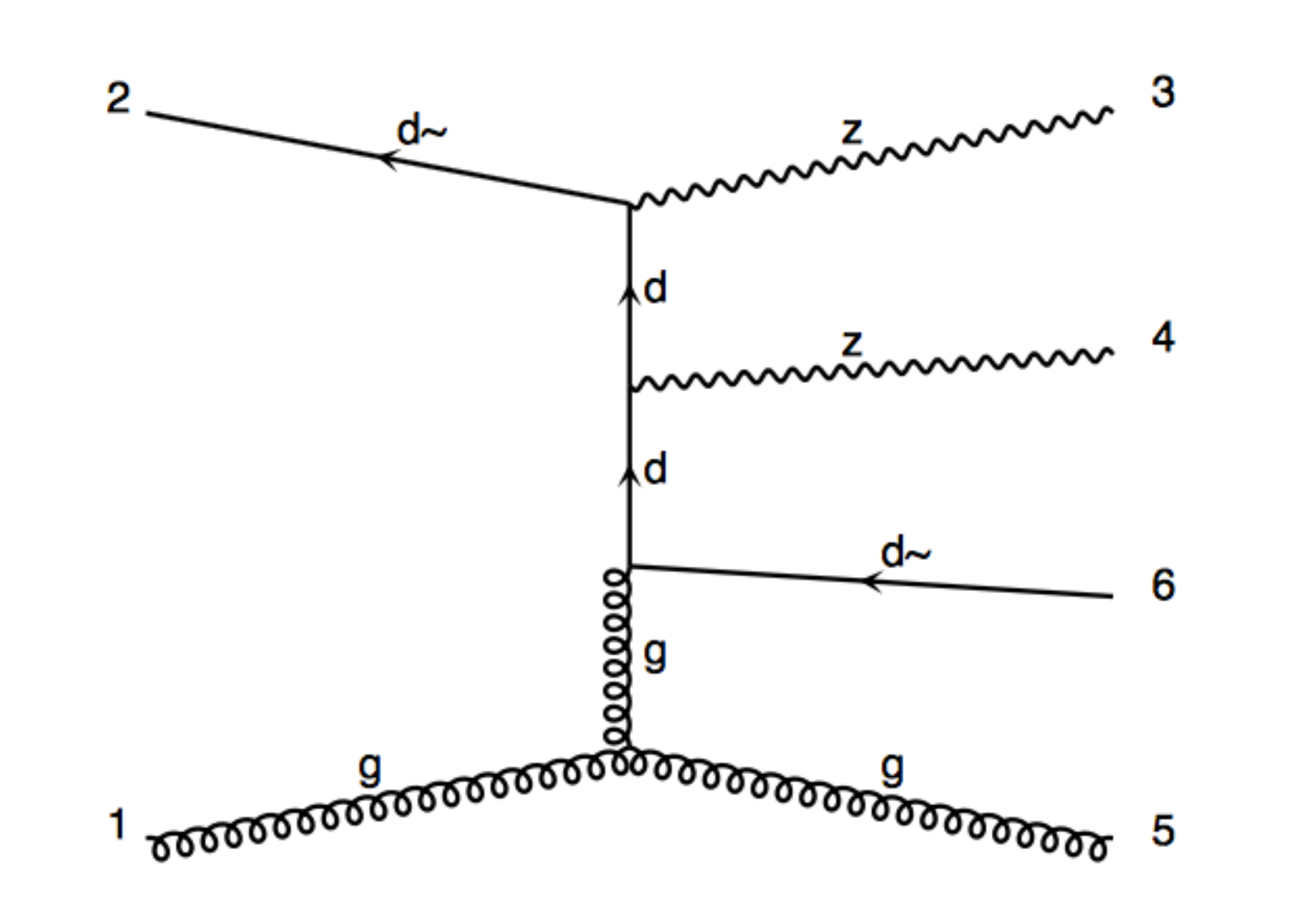} 	
	\includegraphics[width=.30\textwidth,trim=0 0 0 0,clip]{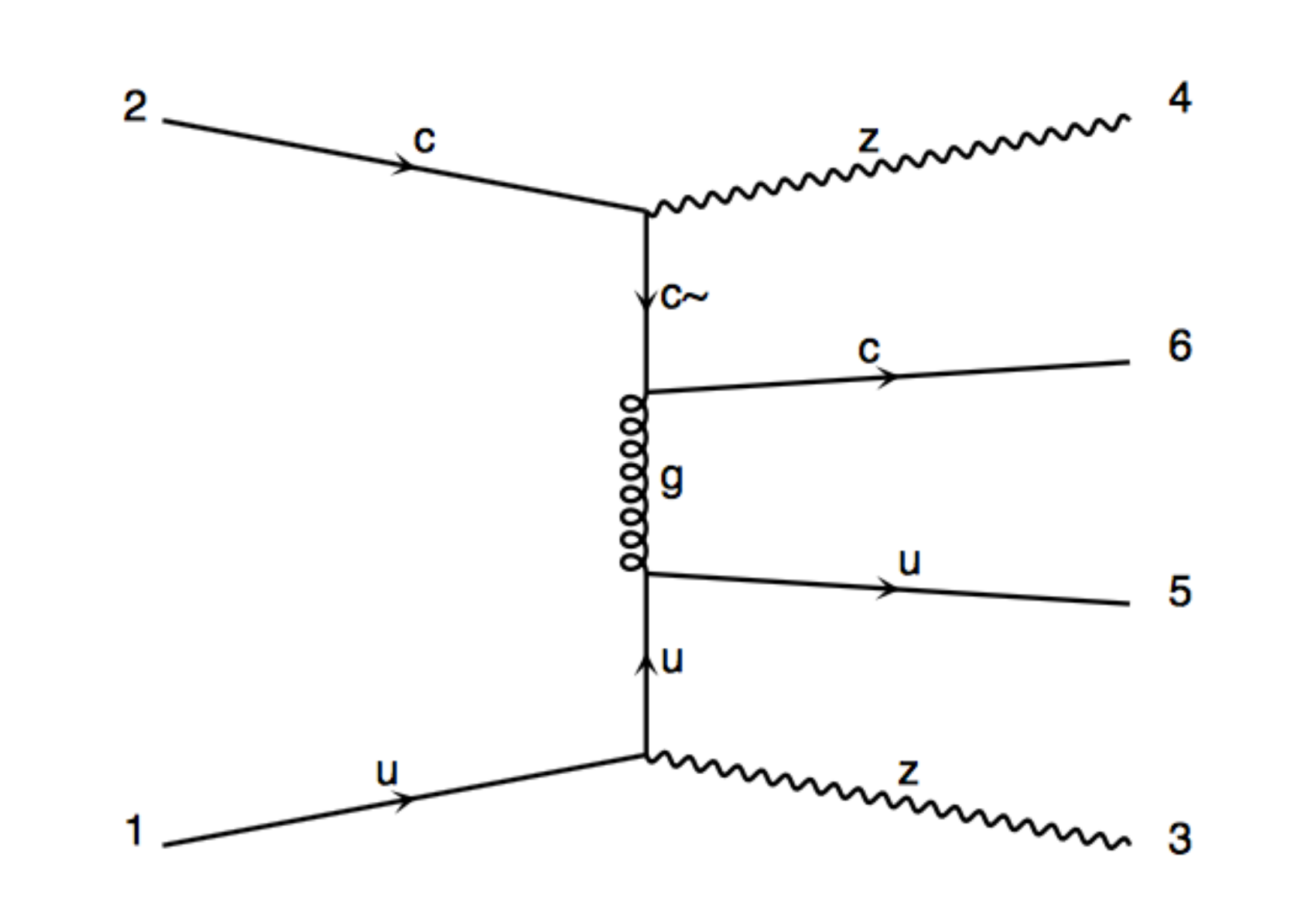} \\
\qquad \qquad	\qquad  d \qquad \qquad \qquad \qquad  \qquad \qquad e  \qquad \qquad \qquad \qquad  \qquad \qquad f \qquad \qquad \qquad \qquad

	\caption{Examples of Feynman diagrams for the different processes of  production of a pair of ZZ bosons  in association with 0, 1 and 2 jets at tree-level.\label{fig:1}
	}	
\end{figure}

At leading order, the process pp $\rightarrow$ ZZ is generated exclusively by the quark-antiquark annihilation($t$-channel) with exchange of quark, a sample is shown in figure~\ref{fig:1}a. 

However, the production of ZZ+jet is mainly via these contributions $q\bar{q}$ $\rightarrow$ ZZ$g$ and $q$($\bar{q}$)$g$ $\rightarrow$ ZZ$q$($\bar{q}$) with exchange of quark and antiquark, some samples are displayed in figure~\ref{fig:1} b and c.

 For the production of ZZ associated with two jets, different sub-processes contribute, which can be grouped as following: $gg$ $\rightarrow$ ZZ$q\bar{q}$, $q$($\bar{q}$)$g$ $\rightarrow$ ZZ$gq$($\bar{q}$), $q\bar{q}$ $\rightarrow$ ZZ$gg$ and $qq$($\bar{q}$$\bar{q}$, $qq^{'}$,$q\bar{q^{'}}$ ) $\rightarrow$ $qq$($\bar{q}$ $\bar{q}$, $qq^{'}$,$q\bar{q^{'}}$) with $q$, $\bar{q}$ and gluon are exchanged particles, the corresponding representative diagrams are depicted in figure~\ref{fig:1} d, e and f.
 We note that adding a jet in the final state of ZZ pair production produces new channels.

Now, we make a comparison between ZZ and WW production of~\cite{ref:35}, hence,  we notice two main remarks, the first is that there are less number of diagrams in the production of ZZ, and the second is that WW production include triple gauge boson and photon vertices which do not exist in ZZ production. These two differences are due to the identical Z-bosons in the final state.

\subsection{Contributions at next leading order}   
\label{subsec:2}

The NLO-QCD corrections of all our processes involve one-loop real and virtual emission corrections, what will produce a new contributions and new partonic channels, that is a great number of topologies contribute for each process. For example the gg initiated sub-process appears only at NLO level for the processes pp $\rightarrow$ ZZ, pp $\rightarrow$ ZZj. 

 It is well-known that these sub-corrections involve infrared (IR) and  ultraviolet (UV) divergences. To remove them, we use the dimensional regularization~\cite{ref:36}. The infrared divergences, whether they come from the real or the virtual corrections, disappear using the Catani-Seymour algorithm~\cite{ref:37}, while the ultraviolet divergences of the virtual corrections are cancelled by the renormalization. After all, we will have got a sum which represent the final result with ultraviolet (UV) and infrared (IR) finite, equally independent of the regularization scheme.

\begin{figure}[tbp]
	\centering	
	\includegraphics[width=.30\textwidth,trim=0 0 0 0,clip]{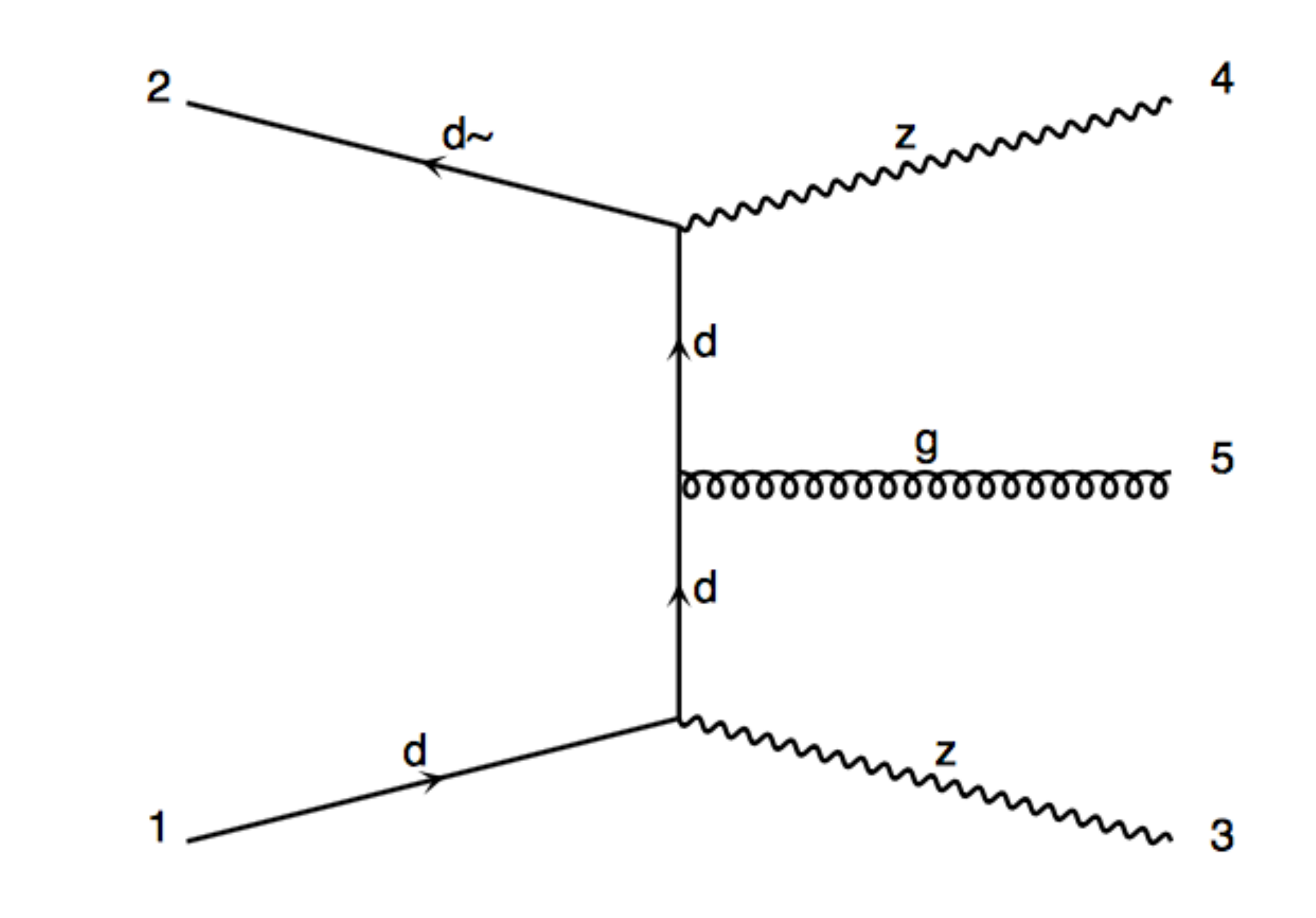}	\includegraphics[width=.30\textwidth,trim=0 0 0 0,clip]{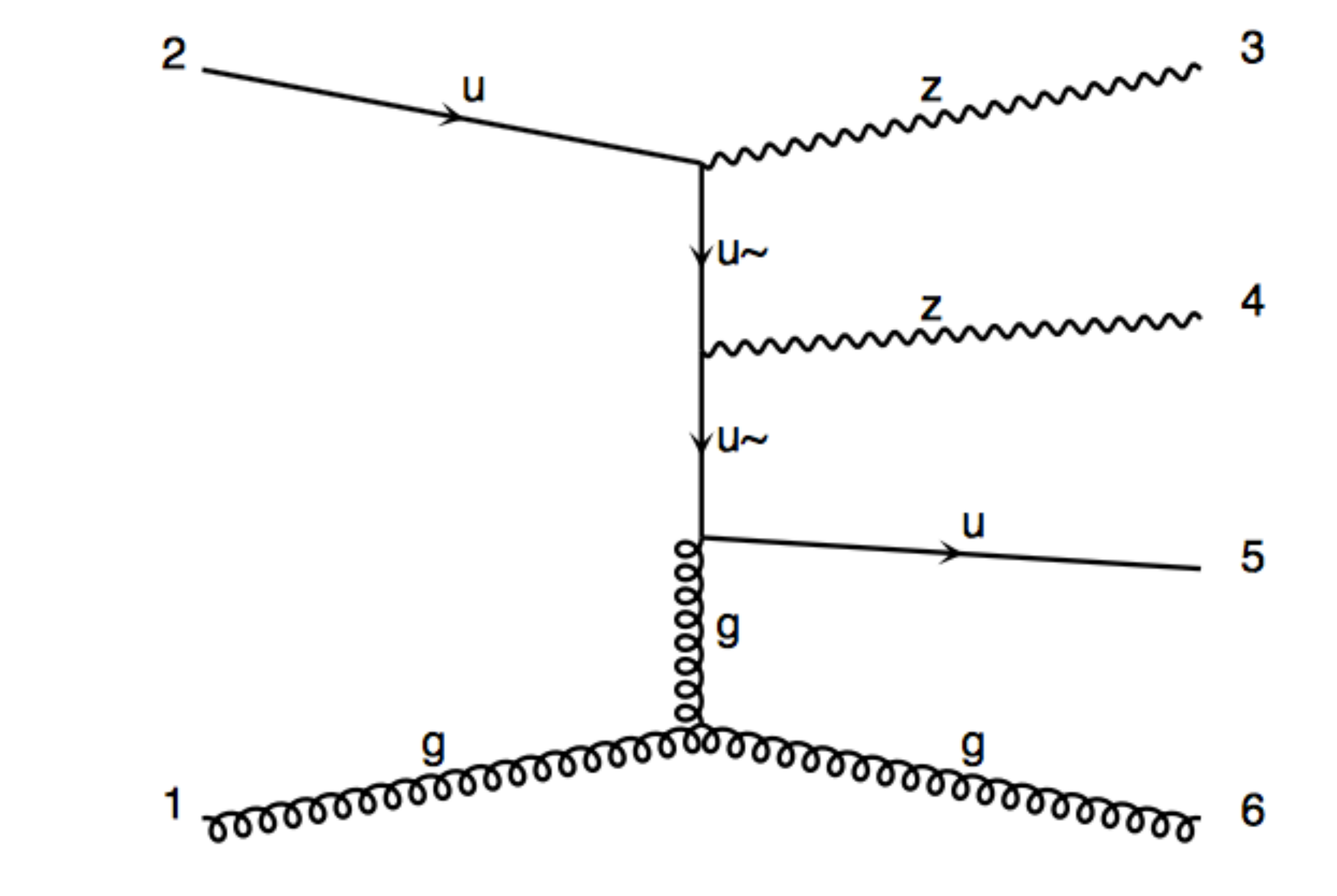} 	
	\includegraphics[width=.30\textwidth,trim=0 0 0 0,clip]{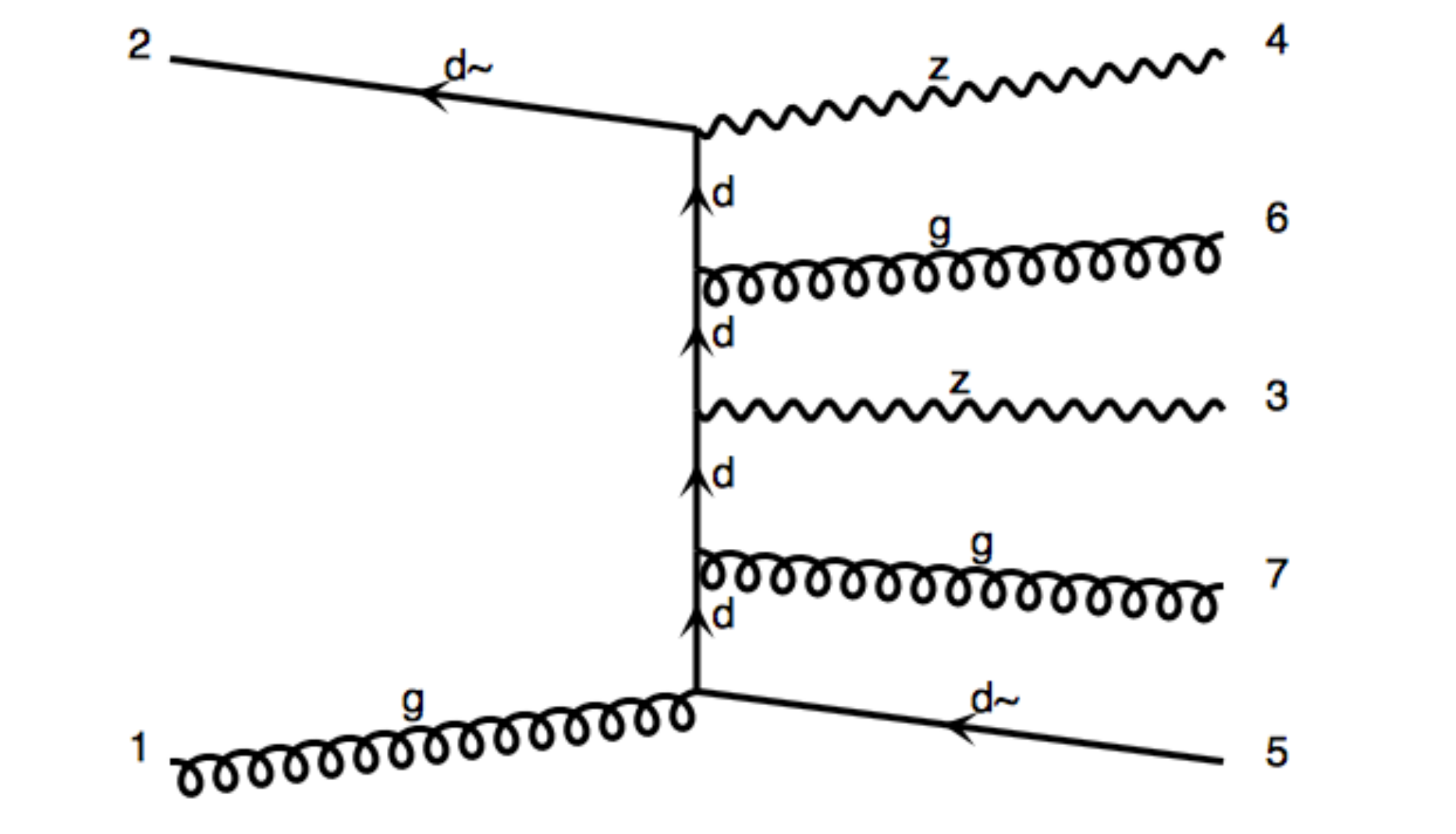} \\
	a \qquad \qquad \qquad \qquad  \qquad \qquad b  \qquad \qquad \qquad \qquad  \qquad \qquad c \qquad \qquad \qquad \qquad
	\includegraphics[width=.30\textwidth,trim=0 0 0 0,clip]{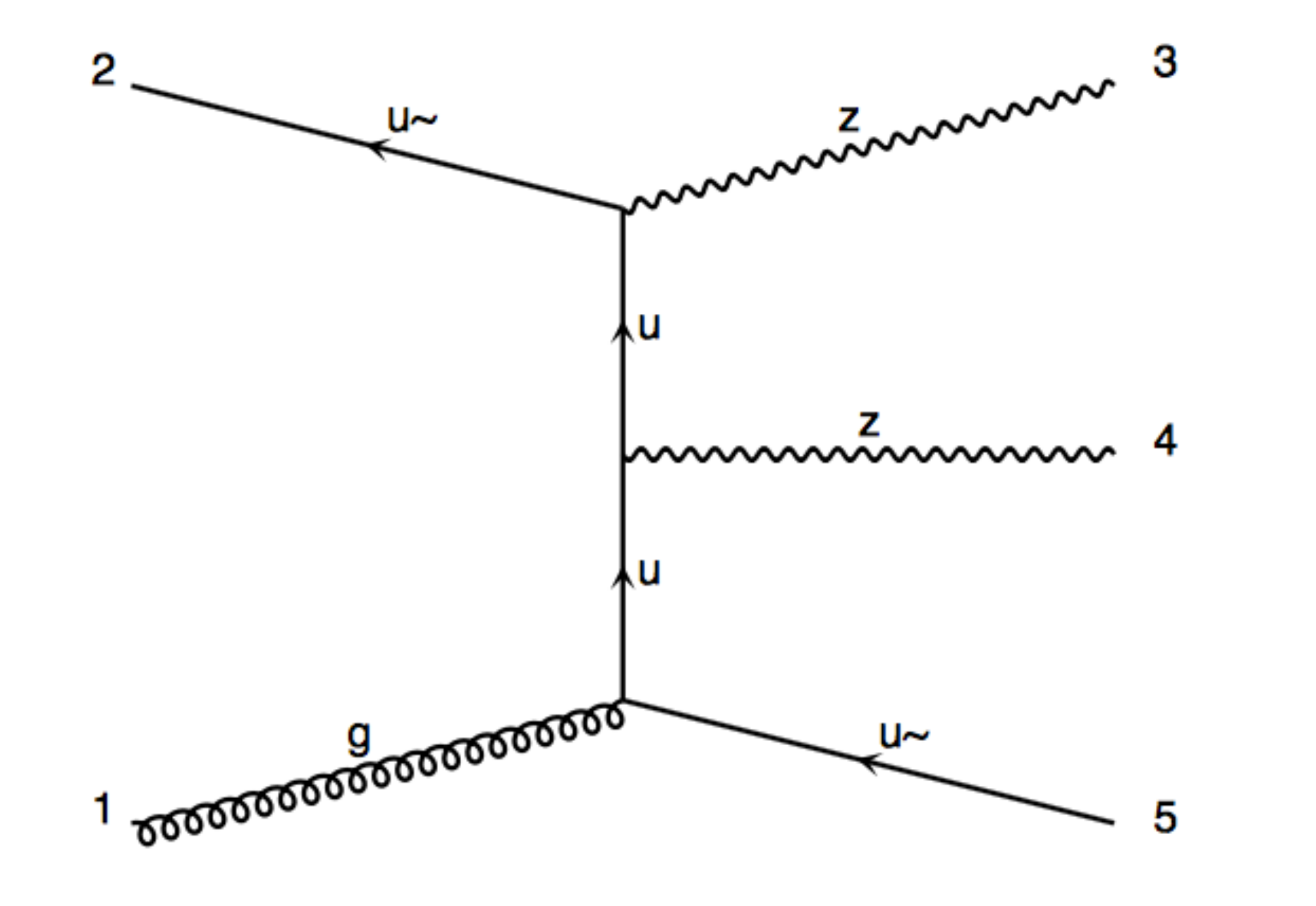}	\includegraphics[width=.30\textwidth,trim=0 0 0 0,clip]{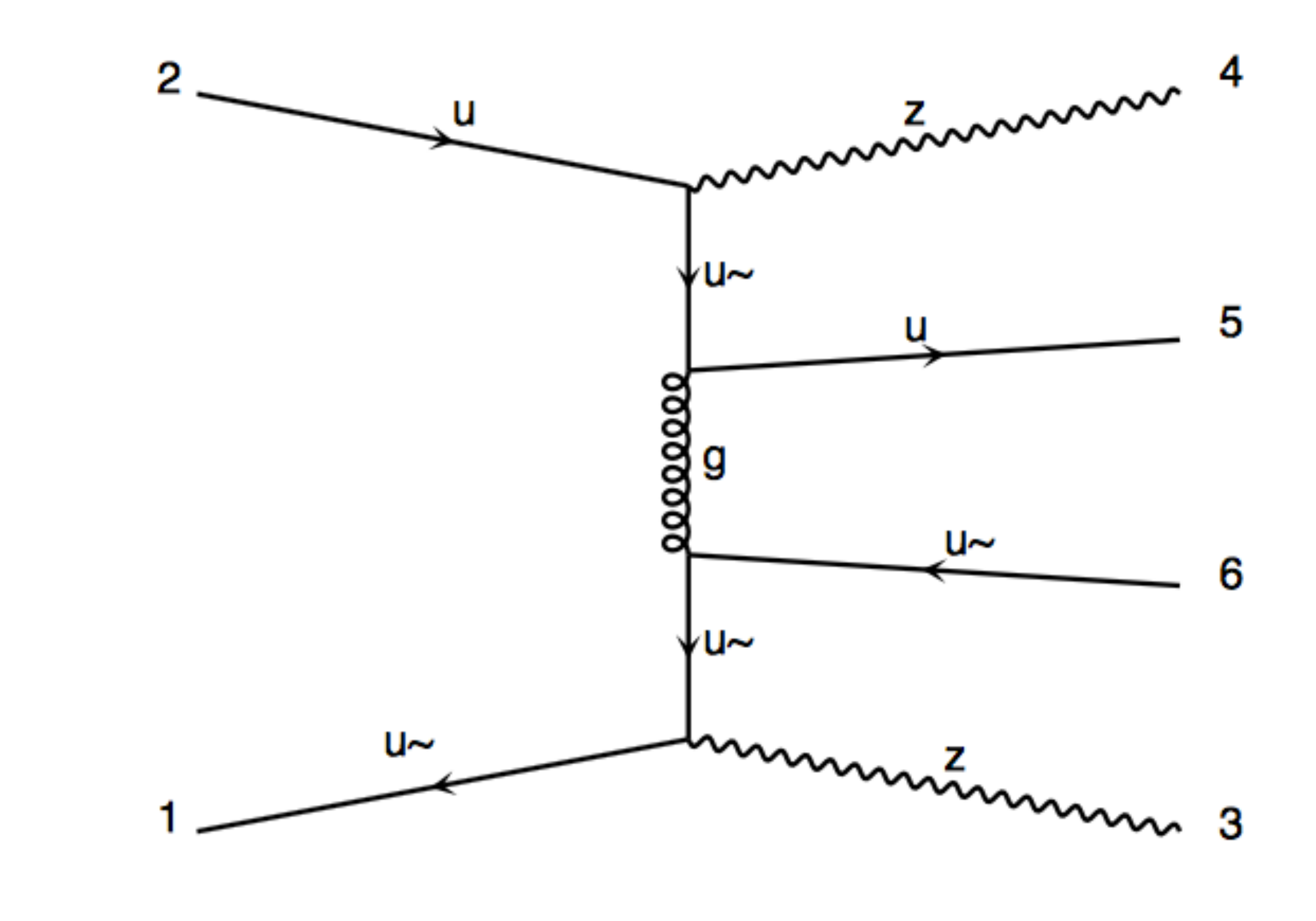} 	
	\includegraphics[width=.30\textwidth,trim=0 0 0 0,clip]{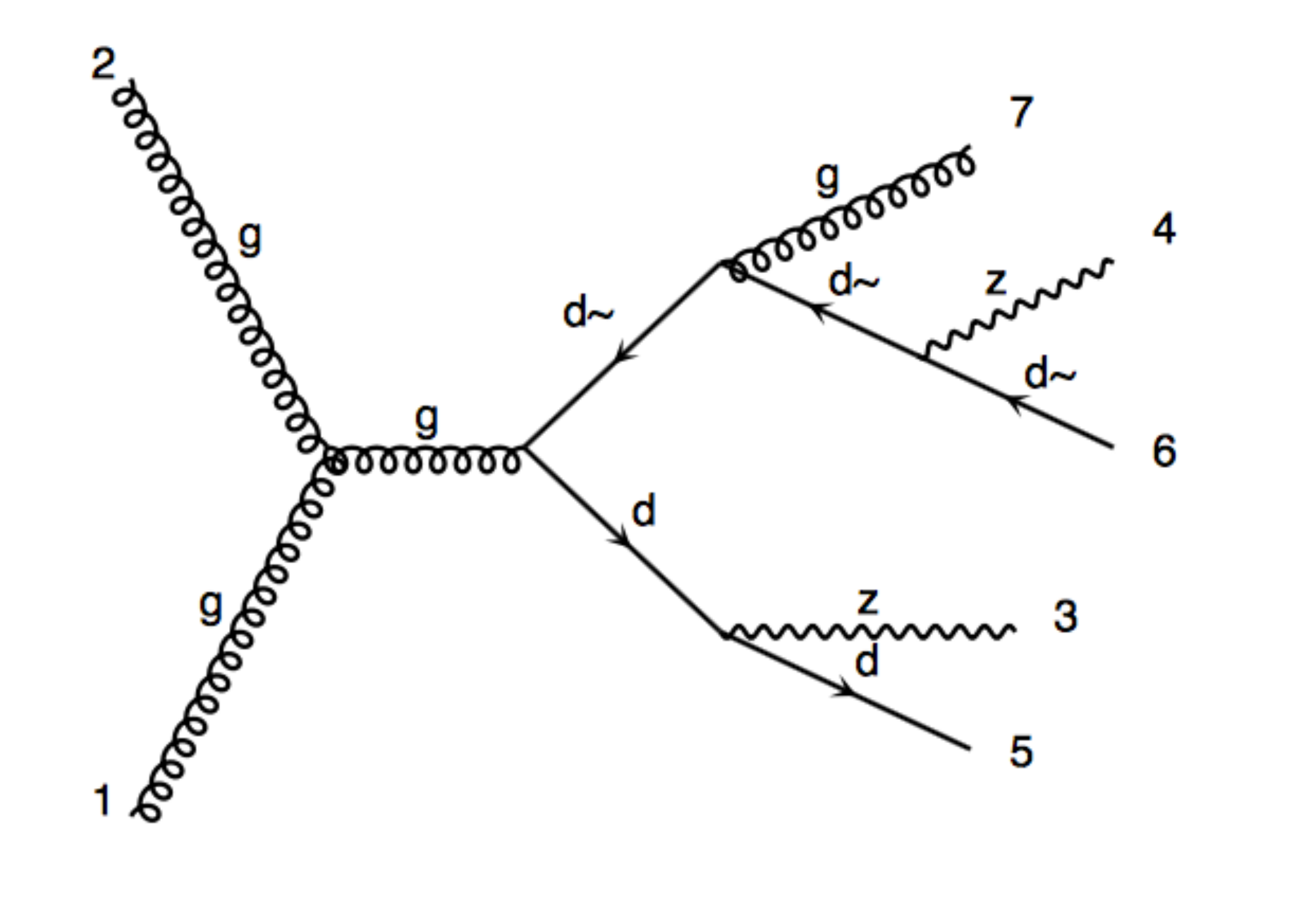} \\
	\qquad \qquad	\qquad  d \qquad \qquad \qquad \qquad  \qquad \qquad e  \qquad \qquad \qquad \qquad  \qquad \qquad f \qquad \qquad \qquad \qquad
	
	\caption{Examples of Feynman diagrams of real corrections for the different processes of  production of a pair of ZZ bosons  in association with 0, 1 and 2 jets. .\label{fig:2}
	}	
\end{figure}

We begin with the real corrections which are the tree level contributions with additional parton (gluons, quarks or anti-quarks) in the final state, that mean, the process p p $\rightarrow$ ZZ will have a third particle in the final state, the same is true for other productions. 
The additional parton can be a gulon obtained only by quark - anti-quark initiated sub-processes for ZZ production as shown in figure~\ref{fig:2} a. On the other hand for the productions of ZZ + j and ZZ + jj, it comes equally from the quark (anti-quark ) gluon channels as represented in figures~\ref{fig:2} b and c.
 Meanwhile, the quark and the anti-quark additional in the final state, via quark (or anti-quark) gluon contributions for ZZ production as illustrated in figure~\ref{fig:2} d. For the processes pp $\rightarrow$ ZZj and pp $\rightarrow$ ZZjj, it is produced by $q\bar{q}$, $qq$, $\bar{q}\bar{q}$, $\bar{q}\bar{q'}$ and $gg$ channels whose some examples are given in figure~\ref{fig:2} e and f.

\begin{figure}[tbp]
	\centering	
	\includegraphics[width=.30\textwidth,trim=0 0 0 0,clip]{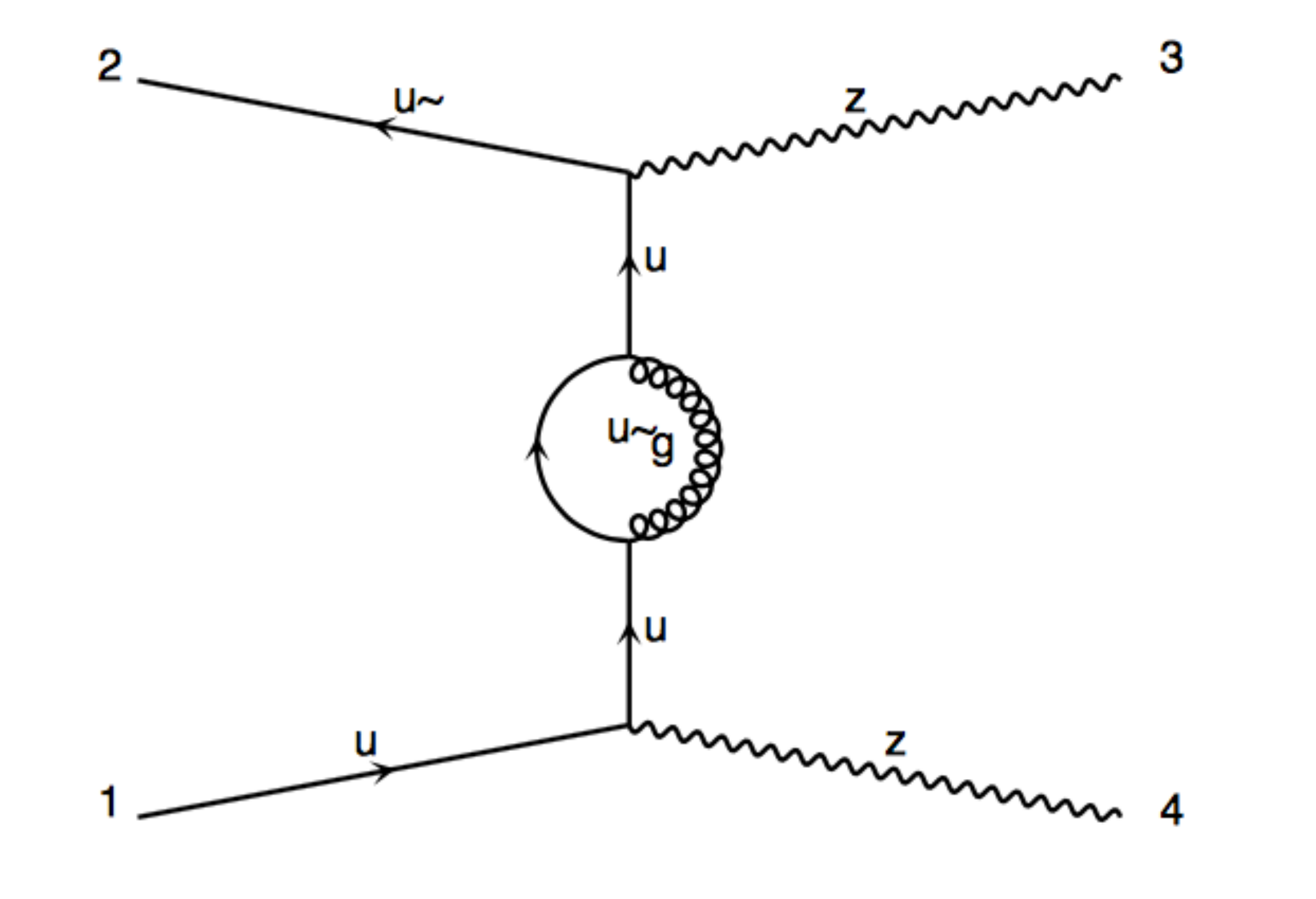}	\includegraphics[width=.30\textwidth,trim=0 0 0 0,clip]{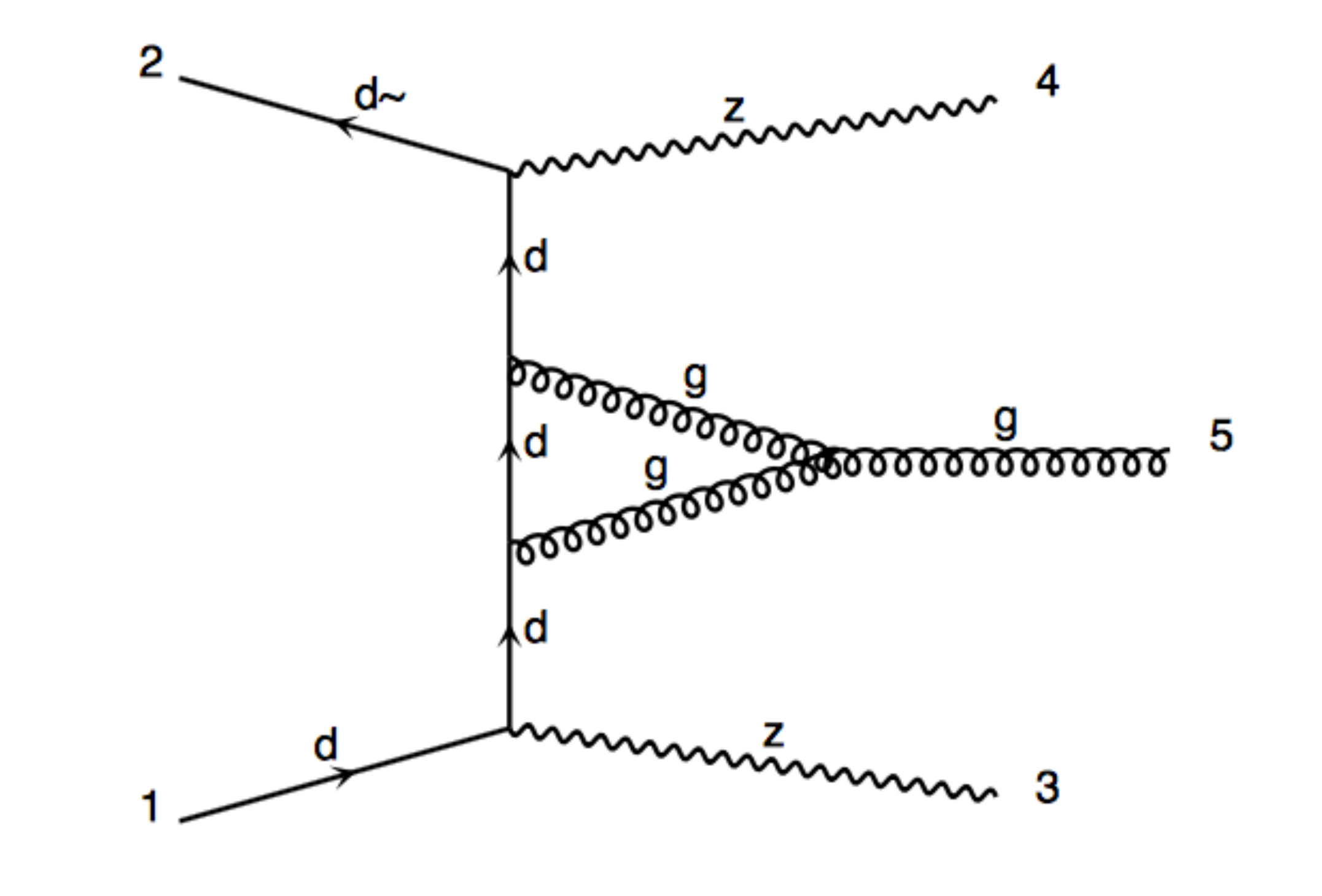} 	
	\includegraphics[width=.30\textwidth,trim=0 0 0 0,clip]{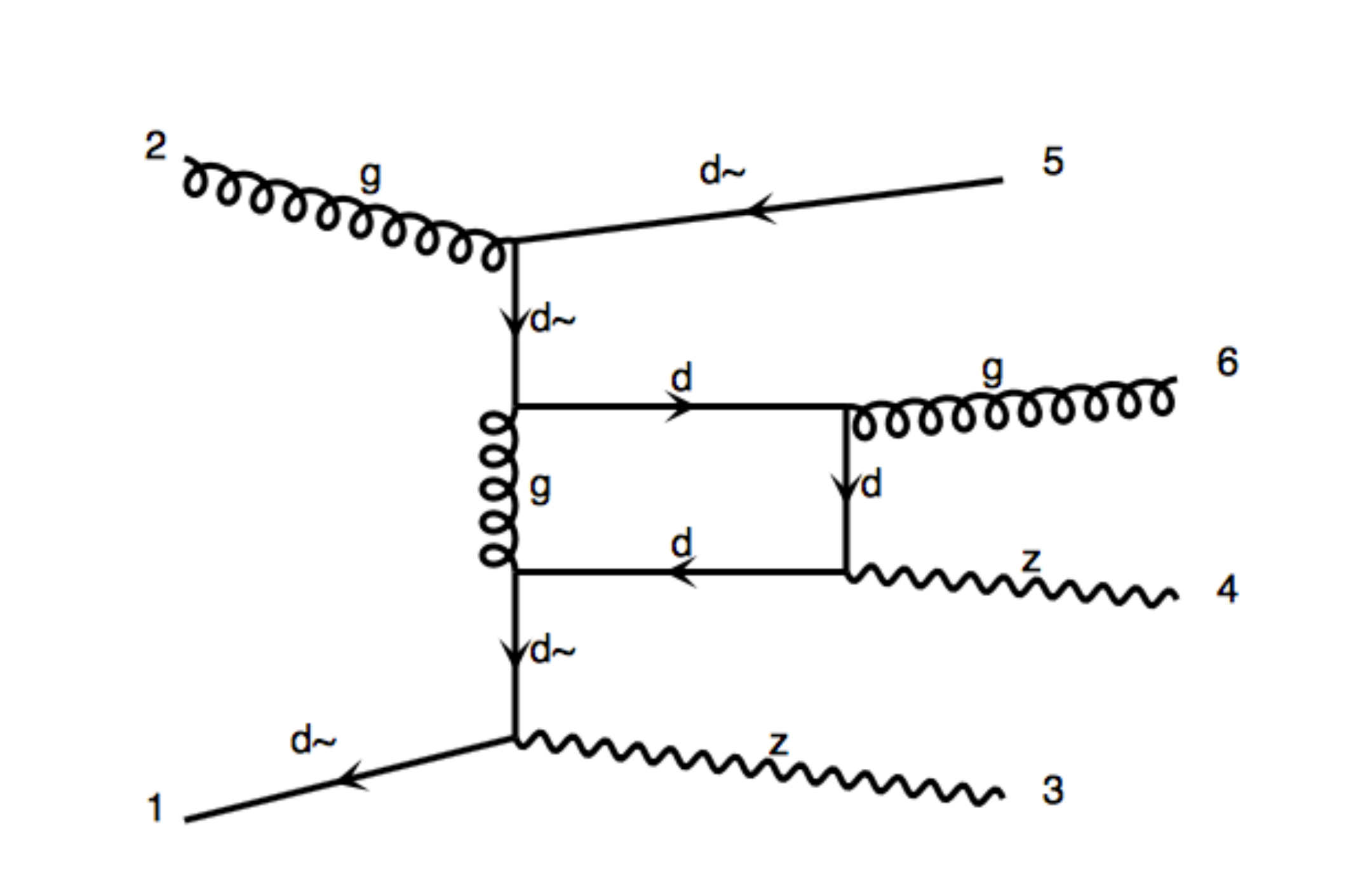} \\
	a \qquad \qquad \qquad \qquad  \qquad \qquad b  \qquad \qquad \qquad \qquad  \qquad \qquad c \qquad \qquad \qquad \qquad

	\caption{Examples of Feynman diagrams of virtual corrections for the different processes of  production of a pair of ZZ bosons  in association with 0, 1 and 2 jets. \label{fig:3}
	}	
\end{figure}    

The virtual corrections involve computing the contributions of the one-loop to the 2 $\rightarrow$ 2 process. There are three types of one-loop diagrams: boxes, triangles, and self-energy which are made up of gluons, quarks, anti-quarks, or a mixture of everything (an example for each is shown in Figure~\ref{fig:3}). For the process p p $\rightarrow$ ZZ, these loops via only from $q\bar{q}$ contributions, contrary to the ZZj and ZZjj production where these loops are included by all the contributions cited in the previous paragraph.

\section{Results  and discussion }
\label{sec:3}

In this section we present the numerical predictions of the integrated cross sections with different cuts in jets transverse momentum and equally the differential distributions of several variables for  processes of~\ref{eq:1}. The events of these calculations are performed in SM at LO and NLO with QCD corrections, at center-of-mass energies of 14 TeV in pp collisions, with Monte Carlo framework implemented in MadGraph5$\_$aMC@NLO at partonic level, after that we have interfaced them with the PYTHIA 8 parton shower and hadronization. In order to reconstruct the events similar to that found at LHC, we use the ATLAS 1604 07773.tcl card~\cite{ref:38} and the fast detector simulation Delphes~\cite{ref:39}.

For all results presented here, we have employed the following SM input parameters.

$G_{\mu}$ = 1.16637.$10^{-5}$ GeV$^{-2}$,  \qquad \qquad $\alpha^{-1}$ = 132.338; 

$M_{W} $ = 80.385 GeV, \qquad \qquad $\Gamma_{w}$ = 2.085 GeV;

$M_{Z} $ = 91.188 GeV, \qquad \qquad $\Gamma_{Z}$ = 2.495 GeV; 

$M_{H} $ = 125 GeV, \qquad \qquad $\Gamma_{H}$ = 0.00407 GeV; 

$M_{t} $ = 173.2 GeV, \qquad \qquad $\Gamma_{t}$ = 1.4426 GeV; 

Throughout the calculation of all our processes, we used NNPDF23 set parton distribution functions as implemented in the LHAPDF setup ~\cite{ref:40}, corresponding to the strong coupling constant $\alpha_{s}$($M_{Z} $) = 0.1137 at LO and $\alpha_{s}$($M_{Z} $) = 0.1251 at NLO. Furthermore, we choose the central values for both the renormalization ($\mu_{R}$) and factorization ($\mu_{F}$ ) scales equal to the pole mass of the Z boson $\mu_{R}$ = $\mu_{F}$ = $M_{Z} $. 

In the meantime, we impose the event selection cuts on the transverse momentum, pseudo-rapidity of the leptons and $\Delta R$ distance (a separation between any possible pair charged leptons).

\begin{equation}
\label{eq:2}
\ |\eta(l)| \leq 2.5, \quad  p_{ T}(l) > 20  GeV  \quad and  \quad \Delta R_{l_{i}l_{j}}= \sqrt{(\eta_{l_{i}} -\eta_{l_{j}})^{2}  + (\phi_{l_{i}} - \phi_{l_{j}})^{2}} > 2
\end{equation}

with $\phi_{l_{i}}$ is the azimuth of lepton i. Thereby, the jets are reconstructed with the anti-kT algorithm~\cite{ref:41} as implemented in the Fastjet program~\cite{ref:42} with a radius parameter set to R = 0.6. Moreover, we require a cut on pseudo-rapidity of jets and the three cuts on the transverse momentum of jets for calculation of the cross sections: 
\begin{equation}
\label{eq:3}
\ |\eta(j)| \leq 2.5, \quad 
p_{ T}(j) > 10  GeV,  \quad    p_{ T}(j) > 100   GeV, \quad p_{ T}(j) > 200   GeV
\end{equation}

\subsection{Total cross sections}

\label{subsec:3.1}

\begin{table}[tbp]
	\centering
	\begin{tabular}{|c|c|c|c|c|}
		\hline
		
			\multicolumn{5}{|c|} {$p^{min}_{T}(j) > 10 GeV$}  \\ 	\hline 
		
		&	$ \sigma_{LO}$[pb]  & $\sigma_{NLO}$[pb]  & $ K =  \frac{\sigma_{NLO}}{\sigma_{LO}}$ & $\sigma_{NLO}^{gg}$     \\ \hline 
		
		$ZZ$     & 9.652 $ \pm$0.026 $^{+5.01\%} _{-5.95\%}$ &14.95$ \pm$0.067 $^{+3.0\%} _{-3.5\%}$        &  1.55  &	1.179$ \pm$0.0009 $^{+21.61\%} _{-17.1\%}$ \\  \hline
		
		$ZZj$  &8.415 $ \pm$0.024 $^{+12.5\%} _{-10.9\%}$& 11.4 $ \pm$0.092  $^{+3.3\%} _{-3.8\%}$    &  1.35  & $ 1.158\pm$0.0 0014$^{+36.7\%} _{-25.1\%}$  \\ \hline 
		
	$ZZjj$    &  5.777 $ \pm$0.013 $^{+22.7\%} _{-17.1\%}$ & 7.041 $ \pm$0.065 $^{+3.1\%} _{-4.0\%}$ &  1.22  & 0.223$ \pm$0.005$^{+38.6\%} _{29.3\%}$  \\ \hline	
		
	\end{tabular} 
		\caption{The LO and NLO total cross section of ZZ pair production with 0, 1 and 2 jets processes at  $\sqrt{s}= 14$ TeV, with a cut on jet transverse momentum $p_{T}(j) > 10$ GeV. }
	\label{tab:1}
\end{table}

\begin{table}[tbp]

	\centering
	\begin{tabular}{|c|c|c|c|c|}
		\hline
		
			\multicolumn{5}{|c|} {$p^{min}_{T}(j) > 100GeV$}    	\\ \hline 
		
		&	$ \sigma_{LO}$[pb]  &$\sigma_{NLO}$[pb]  & $ K =  \frac{\sigma_{NLO}}{\sigma_{LO}}$   &$\sigma_{NLO}^{gg}$    \\  \hline 
		
			$ZZ$     & 9.652 $ \pm$0.026 $^{+5.01\%} _{-5.94\%}$ & 15$ \pm$0.052   $^{+3.1\%} _{-3.6\%}$      &  1.55 & 1.177$ \pm$0.0009 $^{+21.61\%} _{-17.3\%}$  \\ \hline
		
			$ZZj$   & 0.727$ \pm$0.0021$^{+12.2\%}_{-10.3\%}$ &  1.231$ \pm$0.0062$^{+7.8\%}_{-6.8\%}$  &  1.69  &  0.129$ \pm$0.0001$^{+38.1\%}_{-24.7\%}$     \\ \hline 
		
			$ZZjj$   &  0.1996$ \pm$0.00051$^{+26.6\%}_{-19.4\%}$ & 0.2681$ \pm$0.00014$^{+5.9\%}_{-8.3\%}$ &  1.34 & 0.006$ \pm$5.05.10$^{-5}$$^{+39.9\%}_{-32.1\%}$        \\ \hline	
		
	\end{tabular} 
	\caption{The LO and NLO total cross section of ZZ pair production with 0, 1 and 2 jets processes at  $\sqrt{s}= 14$ TeV, with a cut on jet transverse momentum $p_{T}(j) > 100$ GeV. }
\label{tab:2}
\end{table}

\begin{table}[tbp]
	
	\centering
	\begin{tabular}{|c|c|c|c|c|}
		\hline
		
			\multicolumn{5}{|c|} {$p^{min}_{T}(j) > 200 GeV$}    \\	\hline 
		
		&	$ \sigma_{LO}$[pb]  &$\sigma_{NLO}$[pb]  & $ K =  \frac{\sigma_{NLO}}{\sigma_{LO}}$ & $\sigma_{NLO}^{gg}$     \\   \hline 
		
			$ZZ$      & 9.652 $ \pm$0.026 $^{+5.01\%} _{-5.94\%}$  &14.90$ \pm$0.056   $^{+3.4\%} _{-3.8\%}$      &  1.54 & 1.175$ \pm$0.0009 $^{+21.66\%} _{-17.2\%}$ \\ \hline
		
			$ZZj$   & 0.1537$ \pm$0.00047$^{+14.4\%}_{-11.9\%}$ &  0.304$ \pm$0.0018$^{+10.1\%}_{-8.9\%}$   &  1.98 & 0.03$ \pm$3.2.10$^{-5}$$^{+40.2\%}_{-25.9\%}$       \\ \hline 
		
			$ZZjj$     &  0.0481$ \pm$0.00012$^{+27.5\%}_{-20.1\%}$ &  0.0649$ \pm$0.0003$^{+3.8\%}_{-7.3\%}$ &  1.35 & 0.001$ \pm$1.04.10$^{-5}$$^{+40.6\%}_{-33.1\%}$                  \\ \hline	
		
	\end{tabular} 
		\caption{The LO and NLO total cross section of ZZ pair production with 0, 1 and 2 jets processes at  $\sqrt{s}= 14$ TeV, with a cut on jet transverse momentum $p_{T}(j) > 200$ GeV. }
	\label{tab:3}
\end{table}

In table~\ref{tab:1}, we summarize the numerical results of the on-shell LO and NLO total cross sections, at $\sqrt{s}= 14$ TeV with the transverse momentum of jets $p_{T}(j) > 10$ GeV. The value of the total cross section of each process is the sum of all contributions over phase space. We see from this table that the QCD corrections increase the NLO total cross section by 22$\%$, 49$\%$  and up to 67$\%$  of the leading-order term for the ZZjj, ZZj and ZZ  production respectively. This is as expected because the sub-processes produced by the new channels of the real and virtual corrections have been taken into account, which have an impact no-negligible on the full cross section. However, the inclusion of the $q$($\bar{q}$)$g$ in the processes has a sizeable effect in our results, it represents between 45$\%$ to 70$\%$ of total cross section. Contrary to the cross section given by the $gg$ fusion which is smallest, it can represent at most 9$\%$ of total cross section. 

In view of the pairs of ZZ recede against the jets emitted in the final state, accordingly the total cross section decreases when adding a jet to the final state as indicated in the table, since it is clearly seen that the total cross section of the production ZZ without jets is the largest. It is the same for the factor K corresponding to the ratio of the predictions $\sigma_{NLO}/ \sigma_{LO}$ which represent the corrections width, therefore the process pp $\rightarrow$ ZZ has the large value of the factor K equal to 1.55.

Indeed, when we make a comparison of our  theoretical predictions of ZZ production with those given by the ref~\cite{ref:15}, we obtain an overall  good agreements. In the same way, our ZZj predictions are approximatively in agreements with those of the ref ~\cite{ref:27}, the small differences are given by the choice of cuts and PDFs.
We now compare our results with those for experimental ZZ measurements given by the LHC carried out at $\sqrt{s}= 13$ TeV in pp collisions.   
The total cross section at the ATLAS Collaborations~\cite{ref:7} corresponding to an integrated luminosity of 36.1 fb$^{-1}$ using data collected during 2015 and 2016 is:

$\sigma^{ATLAS}_{p p \rightarrow ZZ}$= 17.8 $\pm$ 1.0 (stat) $\pm$ 0.7 (syst) $\pm$ 0.4 (lumi) pb,

when the  total cross section given by the CMS collaboration~\cite{ref:10} using the data sample corresponds to an integrated luminosity of 35.9 fb$^{-1}$ recorded in 2016 is:  

$\sigma^{CMS}_{p p \rightarrow ZZ} $  = 17.2 $\pm$  0.5 (stat) $\pm$ 0.7 (syst) $\pm$ 0.4 (theo) $\pm$ 0.4 (lumi) pb

Hence, we find a good agreement between our theoretical predictions and the experimental results. Both the ATLAS and CMS results are compatible with our NLO predictions. 

We report in tables~\ref{tab:1},~\ref{tab:2} and~\ref{tab:3} the dependence of the total cross sections with transverse momentum of the hardest jet cuts, the imposed cuts are applied only to the events of the processes of ZZ with 1 and 2 jets, it is for that we do not see any change in the computations of process pp$ \rightarrow $ZZ. Although, the LO and NLO total cross sections of ZZj and ZZjj productions are sensitive to these cuts, so we observe that they rapidly decrease with the increment of these cuts. In fact, we find that the selection cuts effects is also illustrated in the variation of K factor. We notice an increase of K with the increase of cut, this means that there is an increase in QCD corrections at large $p_{T}(j) $.

In the same tables, we report the theoretical errors accompanying our implementations, the first error after the numerical value of the total cross section is statistical errors coming form numerical integrations, it is less than 0.5. Then, the scale uncertainty is estimated in procentage is evaluated in the perturbative framework by varying $\mu_{R}$ and $\mu_{F}$ simultaneously between 0.5$M_{Z} $ < $\mu_{R}$, $\mu_{F}$ < 2$M_{Z} $ with the constraint 0.5 < $\mu_{R}$/$\mu_{F}$ < 2, we see clearly that the scale uncertainty reduced at NLO for all processes but that does not apply to the gluon fusion channel, we also note that there is a slight increases in this uncertainty at large $p_{T}(j) $. The last error presented is the PDF uncertainty due to choice of the parton distribution functions (PDFs), which in our case is NNPDF23, it can be amount to 3.8$\%$.

\subsection{Distributions }
\label{subsec:3.2}
We turn to discuss the differential distributions of various kinematical variables, at LO and NLO with QCD corrections for the 14 TeV. Before we studying the kinematical distributions at hadronic level with showering and hadronization, we first present the distributions results at parton level.

\subsubsection{Parton level distributions}
 At partonic level, the MadGraph$\_$aMC@NLO generate the first hard emission for processes ~\ref{eq:1} and compress their informations in files so called Les Houches Event (LHE). At this stage, the results are unphysical but we use it to motivate our analysis and compare it to the realistic predictions given by showering and hadronization at hadronic level.

\begin{figure}[tbp]
	\centering	
	\includegraphics[width=.30\textwidth,trim=0 0 0 0,clip]{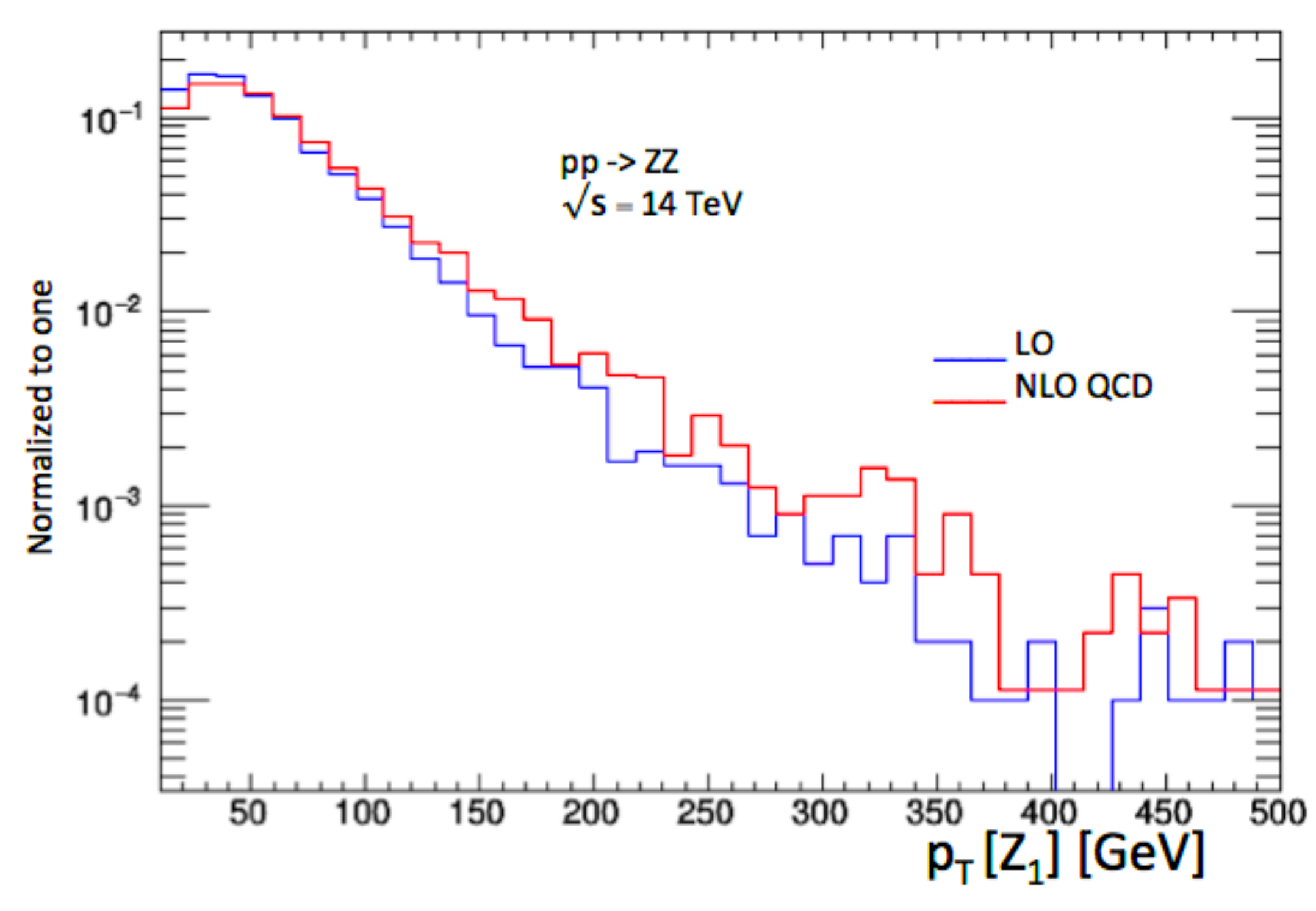}	\includegraphics[width=.30\textwidth,trim=0 0 0 0,clip]{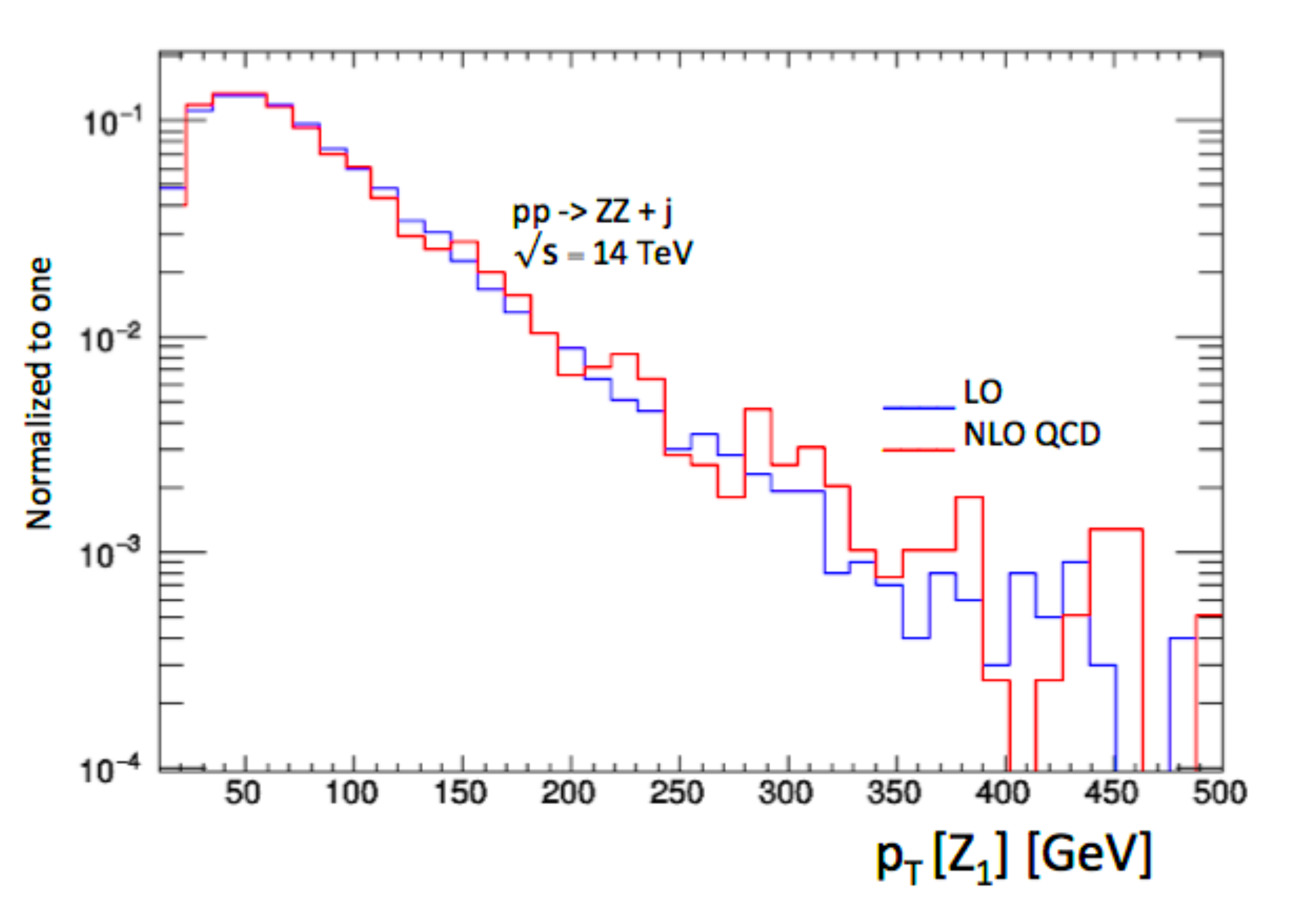} 	
	\includegraphics[width=.30\textwidth,trim=0 0 0 0,clip]{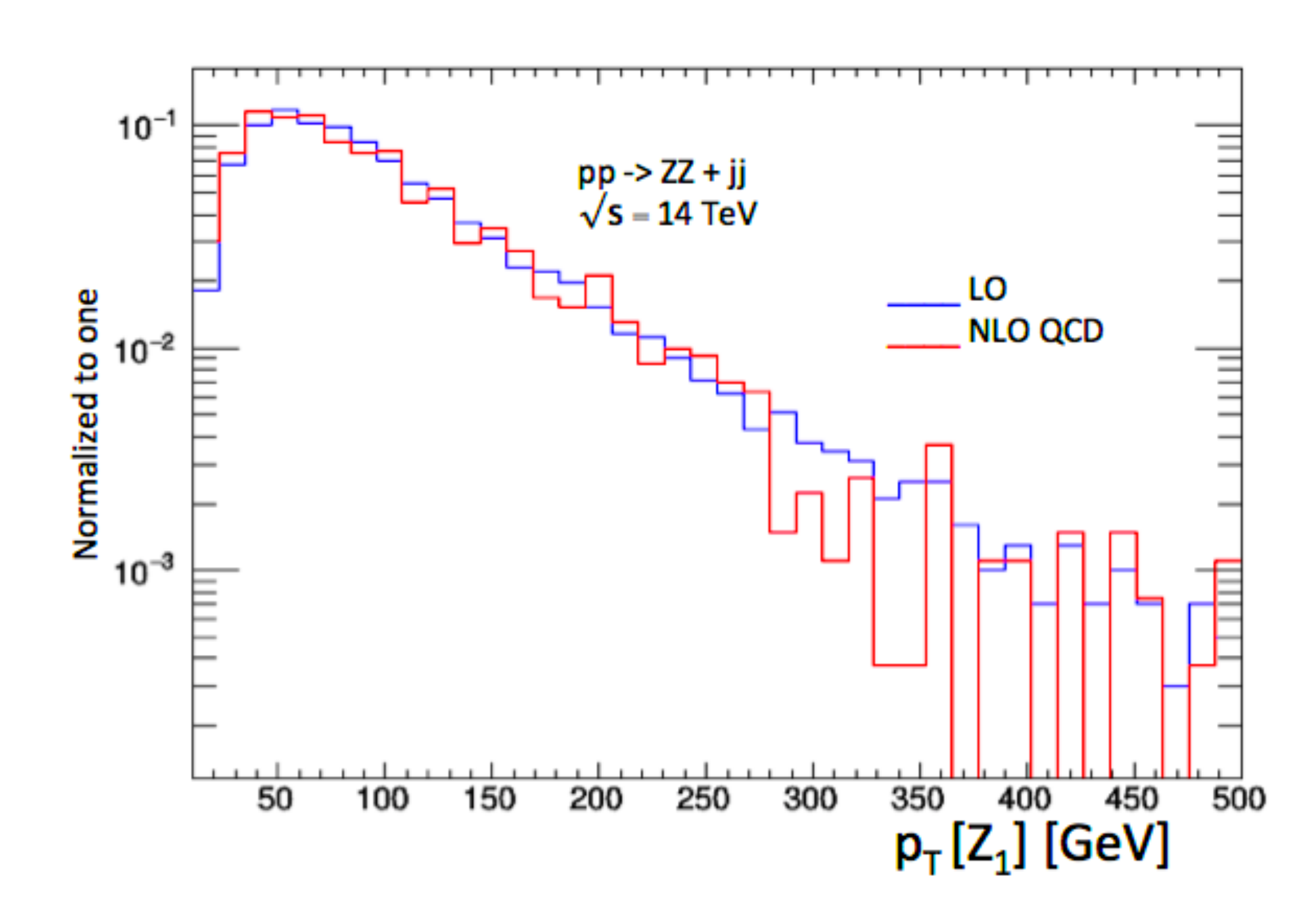} \\

	\caption{The LO and NLO transverse momentum distributions of the  leading Z boson at parton-level for the processes~\ref{eq:1} at 14 TeV.
		\label{fig:4}
	}	
\end{figure}

\begin{figure}[tbp]
	\centering	
	\includegraphics[width=.30\textwidth,trim=0 0 0 0,clip]{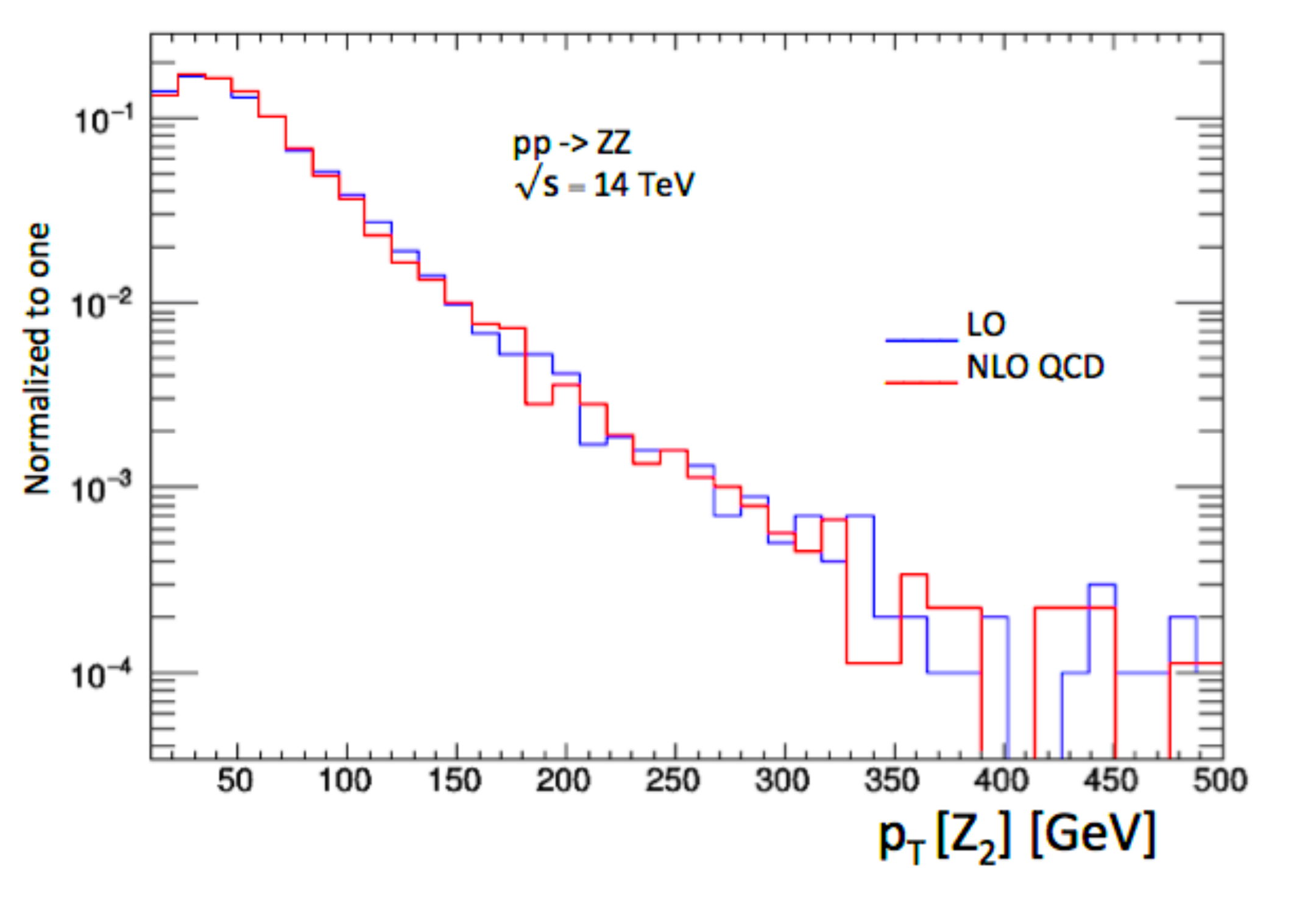}	\includegraphics[width=.30\textwidth,trim=0 0 0 0,clip]{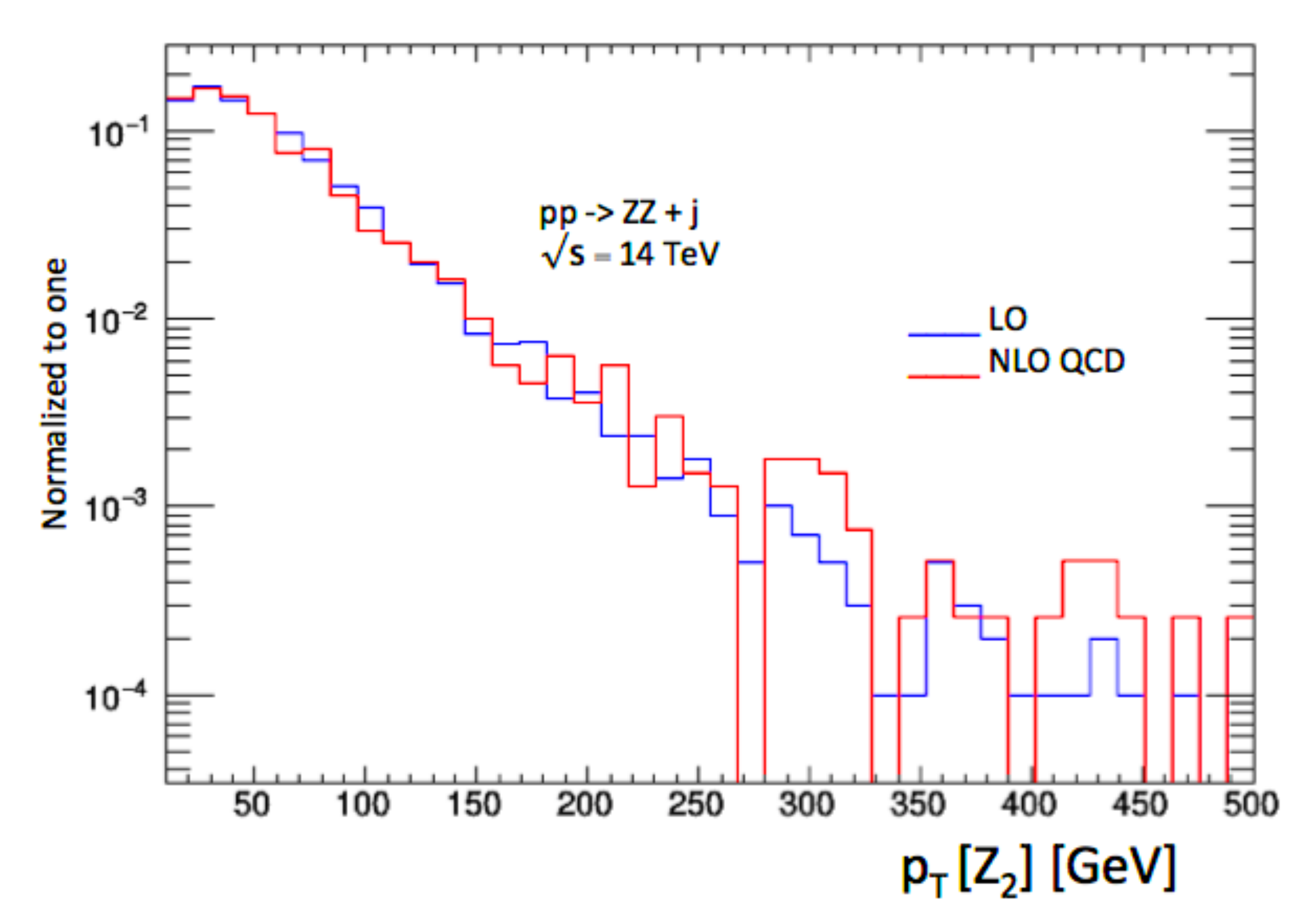} 	
	\includegraphics[width=.30\textwidth,trim=0 0 0 0,clip]{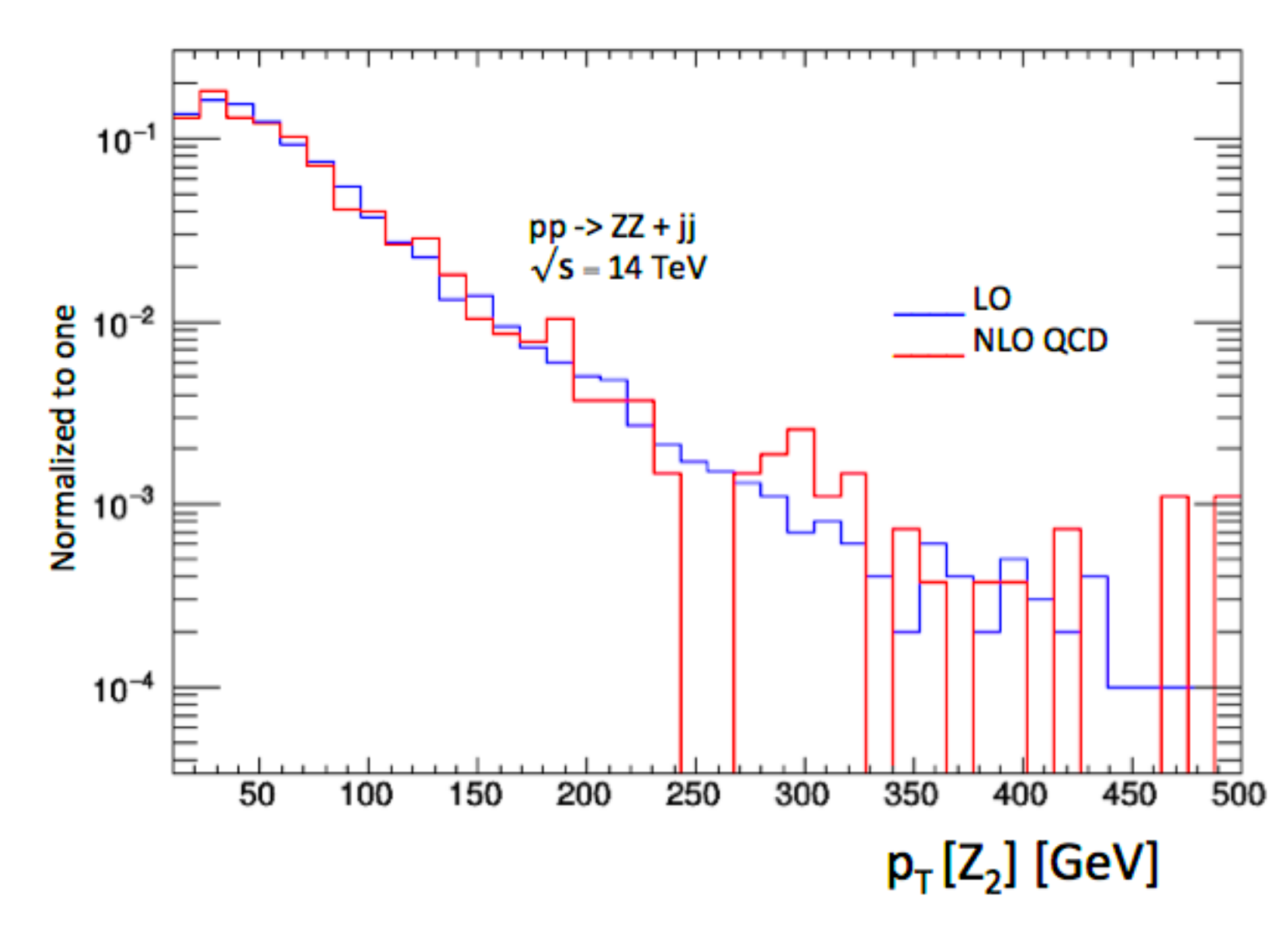} \\

	\caption{The LO and NLO transverse momentum distributions of the subleading Z boson at parton-level for the processes~\ref{eq:1} at 14 TeV.
		\label{fig:5}
	}	
\end{figure}

\begin{figure}[tbp]
	\centering	
	\includegraphics[width=.30\textwidth,trim=0 0 0 0,clip]{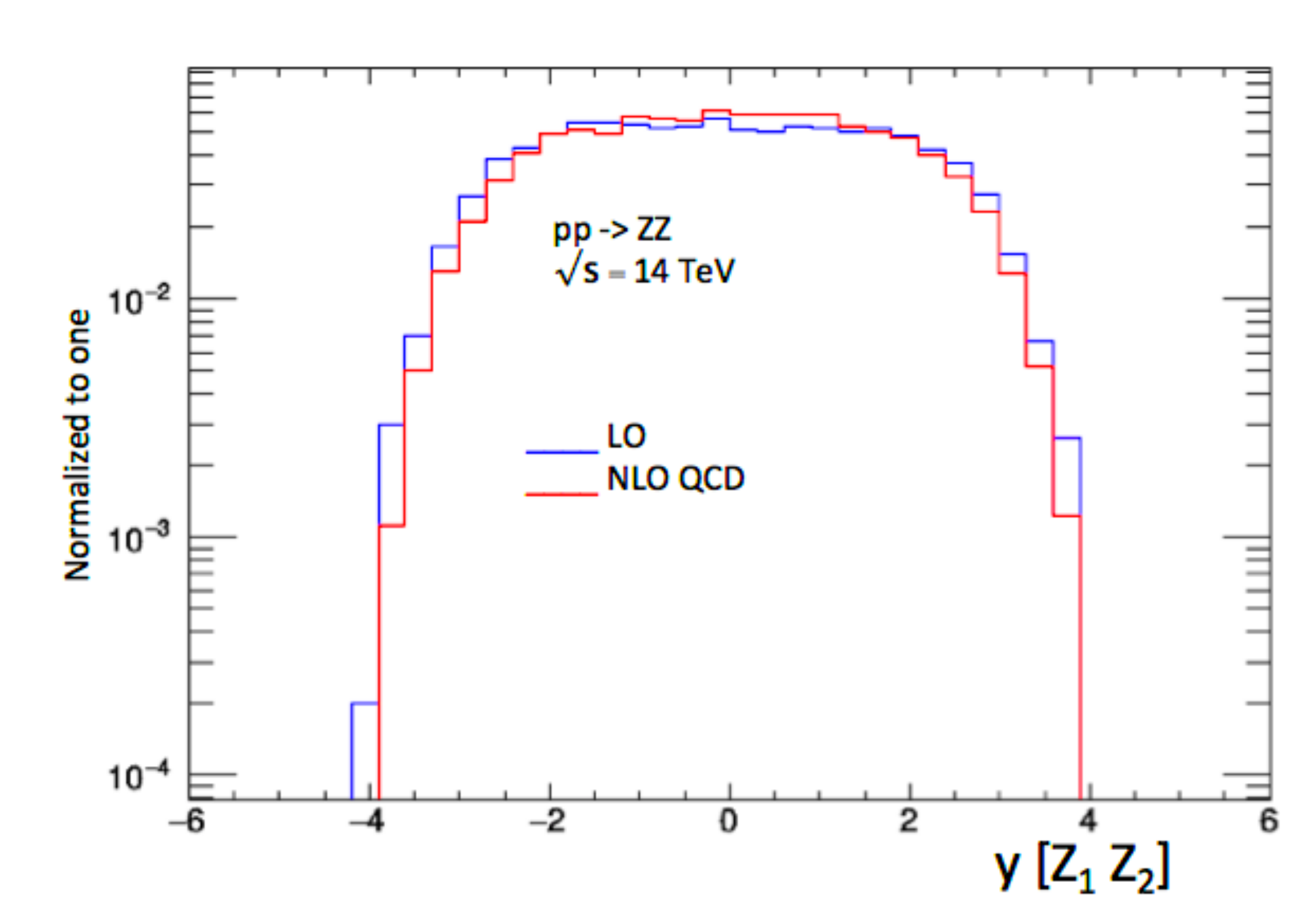}	\includegraphics[width=.30\textwidth,trim=0 0 0 0,clip]{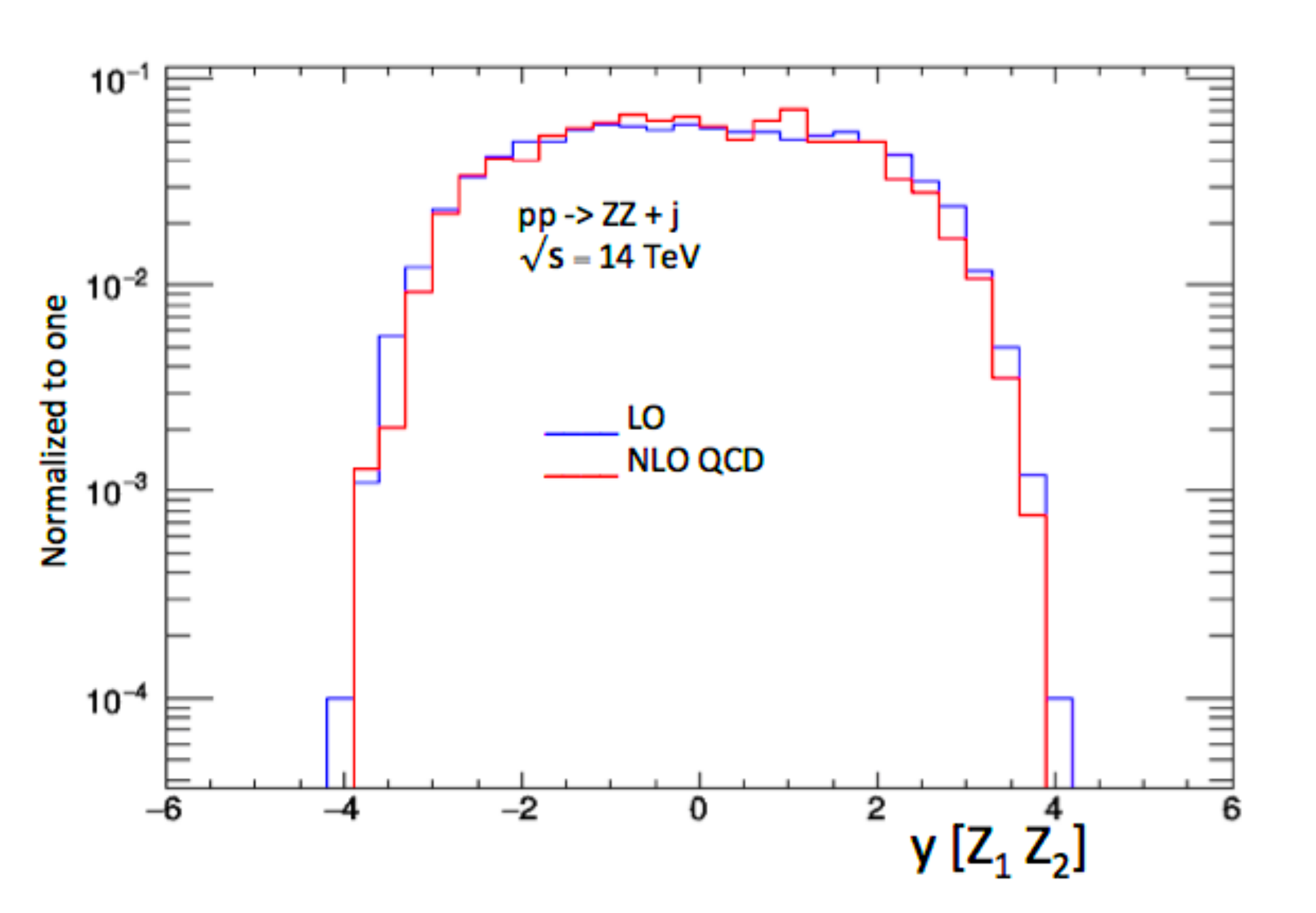} 	
	\includegraphics[width=.30\textwidth,trim=0 0 0 0,clip]{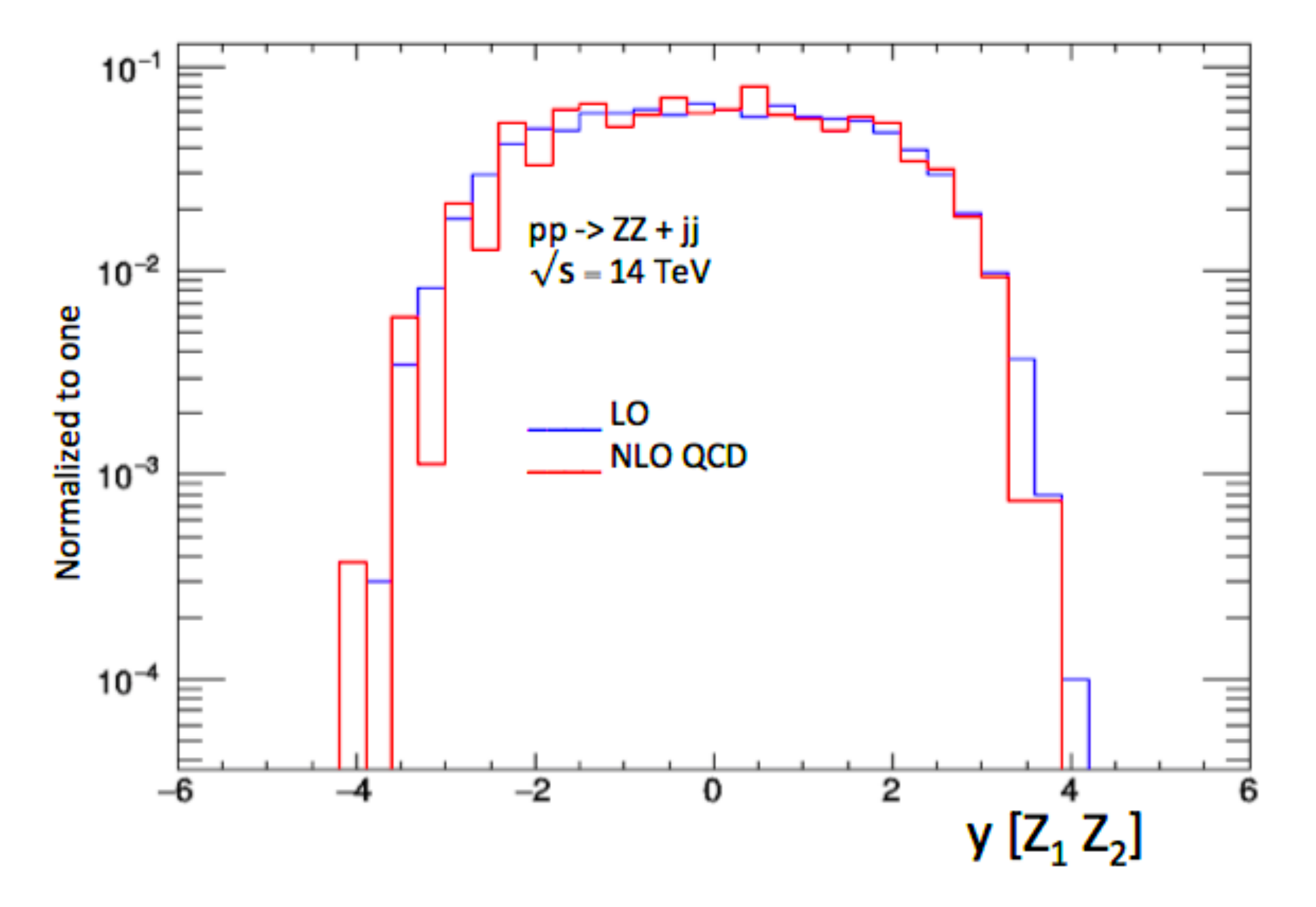} \\

	\caption{Rapidity distributions of Z-boson pair at parton-level for the processes~\ref{eq:1} at 14 TeV.
		\label{fig:6}
	}	
\end{figure}
\begin{figure}[tbp]
	\centering	
	\includegraphics[width=.30\textwidth,trim=0 0 0 0,clip]{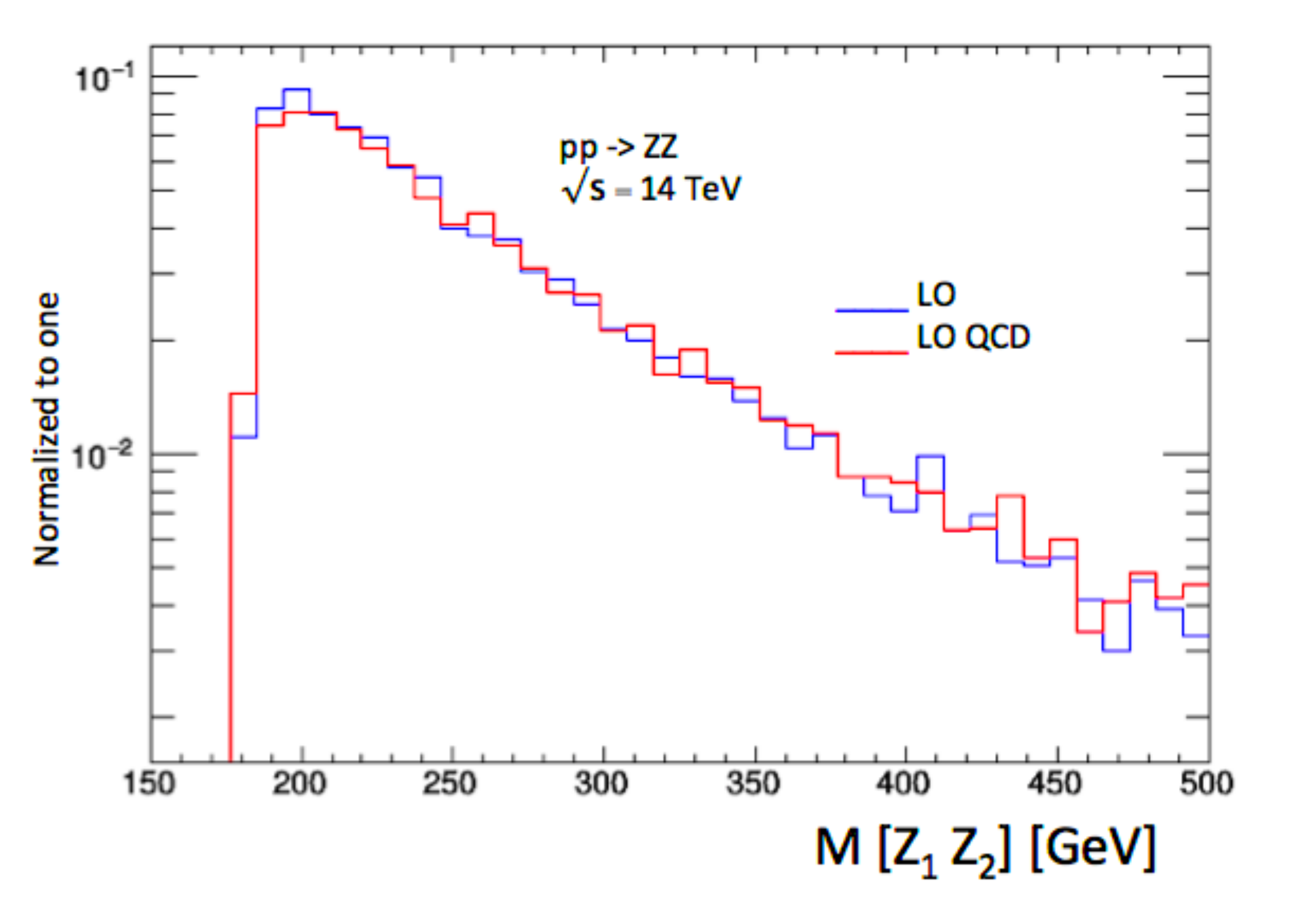}	\includegraphics[width=.30\textwidth,trim=0 0 0 0,clip]{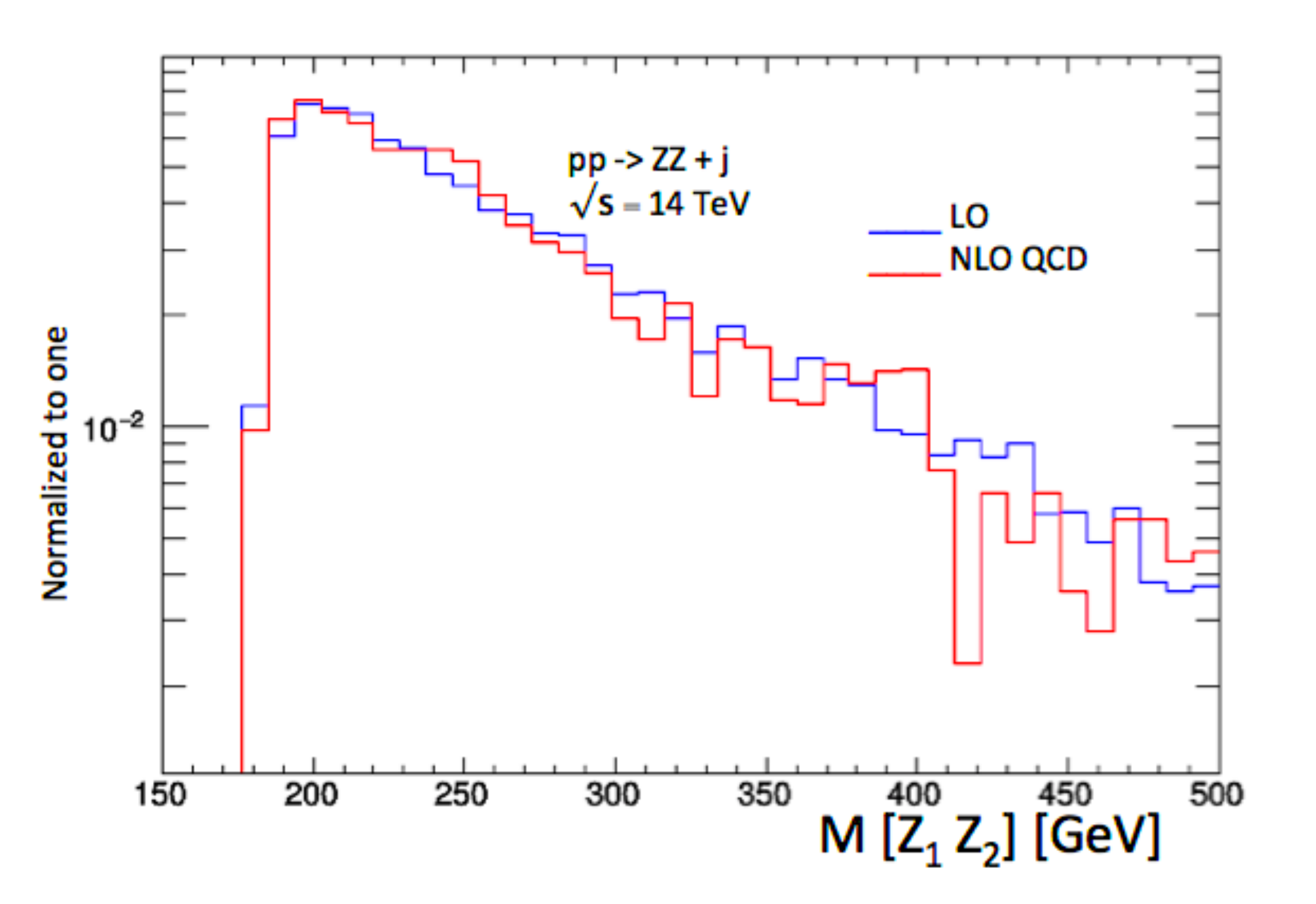} 	
	\includegraphics[width=.30\textwidth,trim=0 0 0 0,clip]{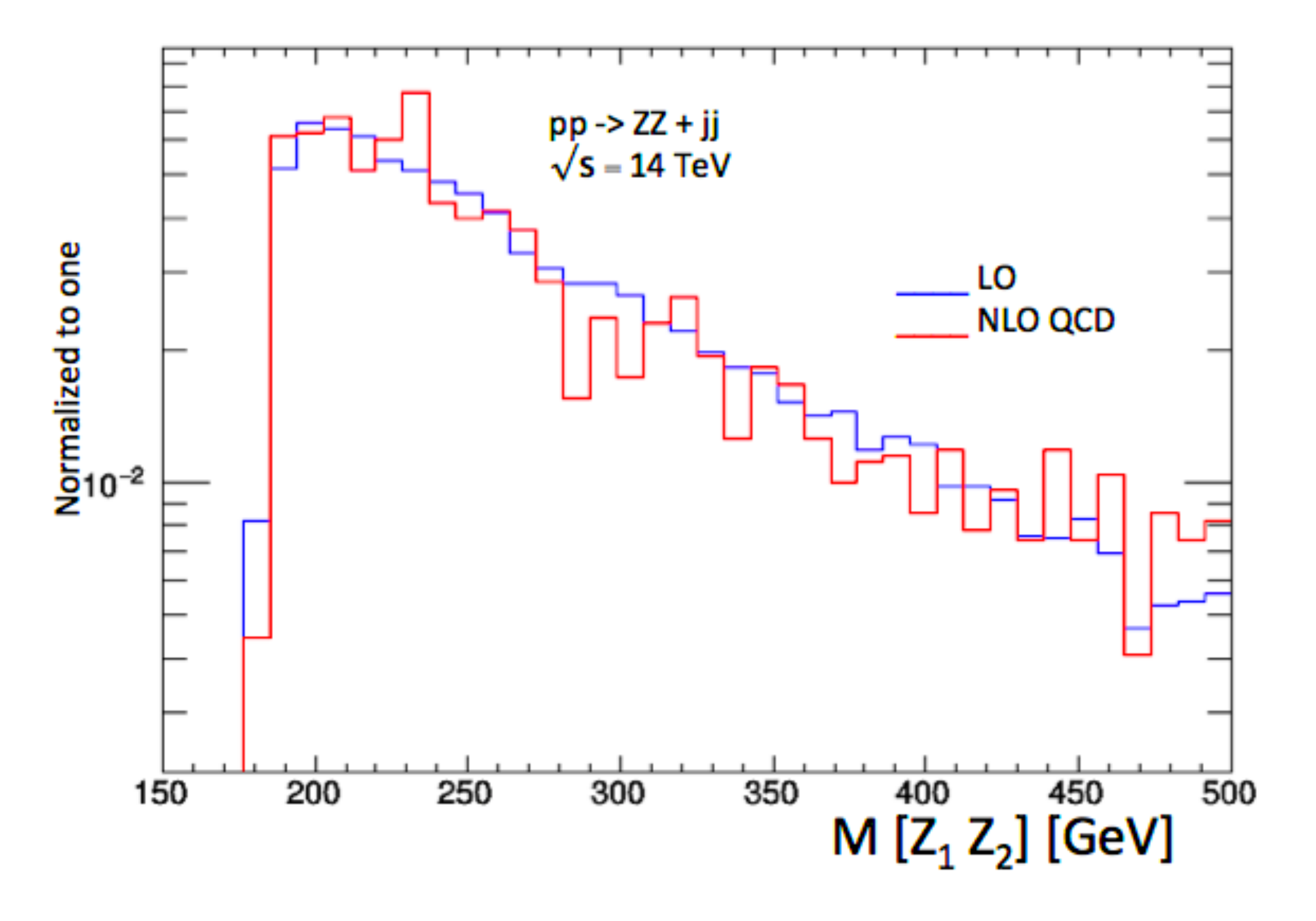} \\

	\caption{ZZ invariant mass distribution at parton-level for the processes~\ref{eq:1} at 14 TeV.
		\label{fig:7}
	}	
\end{figure}

We begin with transverse momentum distributions of the hardest Z boson corresponding to each process, as shown in Figure~\ref{fig:4}. We see that both the LO and NLO QCD put on the transverse momentum distributions peak at 60 GeV for all processes, after that the distributions decrease gradually for the other transverse momentum values. The effect of the QCD corrections are much larger in the $p_{T} $ above 80 GeV for ZZ production and 150 GeV for ZZ production associated with one and two jets. The situation is the same for transverse momentum distributions of the subleading Z boson except that the maximum is at 30 GeV, as illustrated in Figure~\ref{fig:5}, this prediction still agrees well with  those presented in~\cite{ref:8}.

Figure~\ref{fig:6} describes the distribution in the rapidity separation of the two Z bosons, for all processes the interval of $y[Z_{1} Z_{2}]$ is [-4, +4] but the high values of the distributions corresponding to central interval of the rapidity -1.5 < $y[Z_{1} Z_{2}]$ < 1.5.

We now pass to study another interesting distribution is the invariant mass of the ZZ presented in Figure~\ref{fig:7}, we find two region indicated in the curve, thus the distribution increases from M= 2M$_{Z} $ until it will have its maximum in the vicinity of M$_{ZZ}$ $\approx$ 210 GeV, after that it decreases progressively with incerement of M$_{ZZ}$.

\subsubsection{Parton shower distributions }

\begin{figure}[tbp]
	\centering	
	\includegraphics[width=.30\textwidth,trim=0 0 0 0,clip]{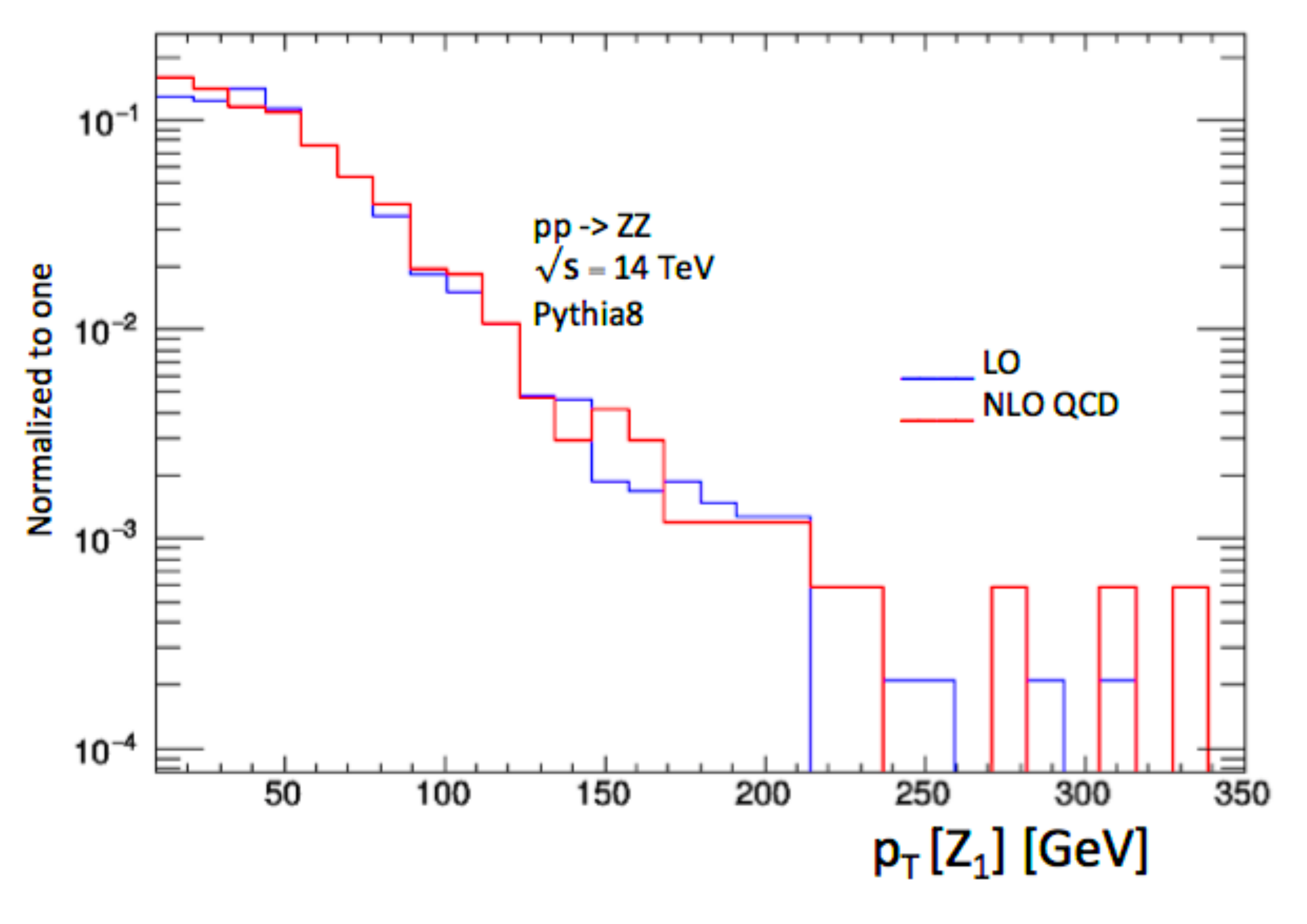}	\includegraphics[width=.30\textwidth,trim=0 0 0 0,clip]{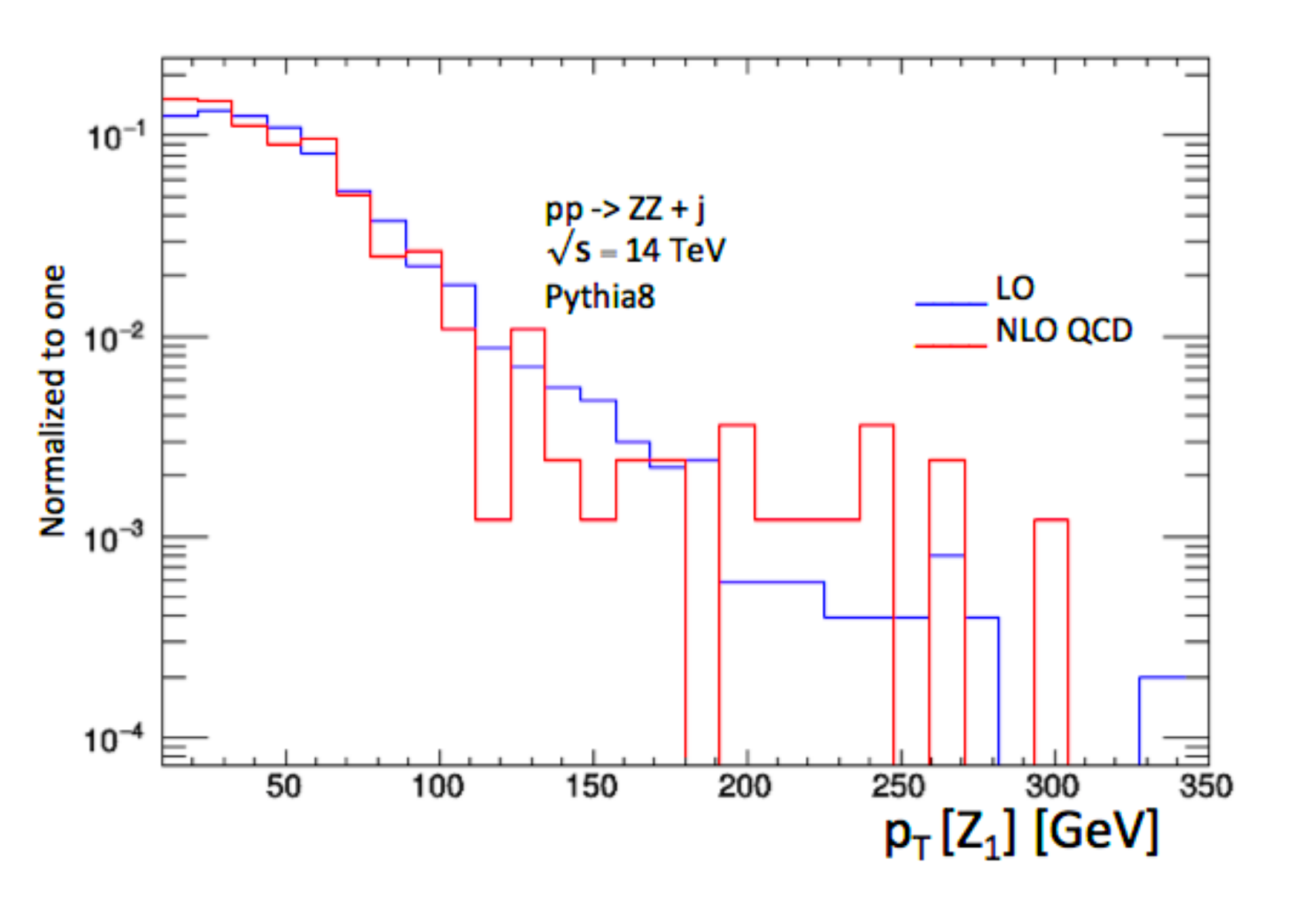} 	
	\includegraphics[width=.30\textwidth,trim=0 0 0 0,clip]{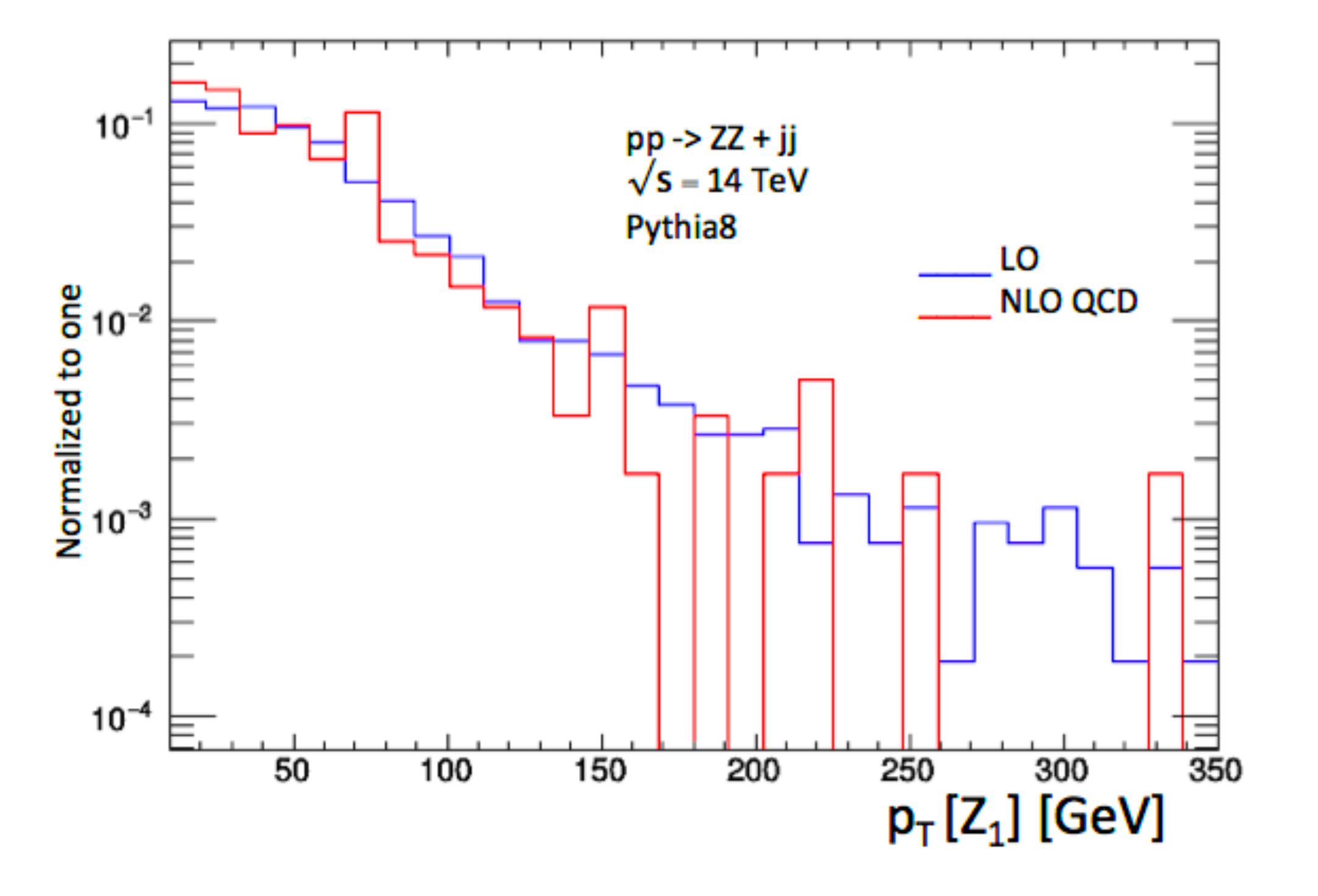} \\

	\caption{The LO and NLO transverse momentum distributions of the the leading Z boson reconstructed after showering and hadronization with PYTHIA 8 for the processes~\ref{eq:1} at 14 TeV.
		\label{fig:8}
	}	
\end{figure}

\begin{figure}[tbp]
	\centering	
	\includegraphics[width=.30\textwidth,trim=0 0 0 0,clip]{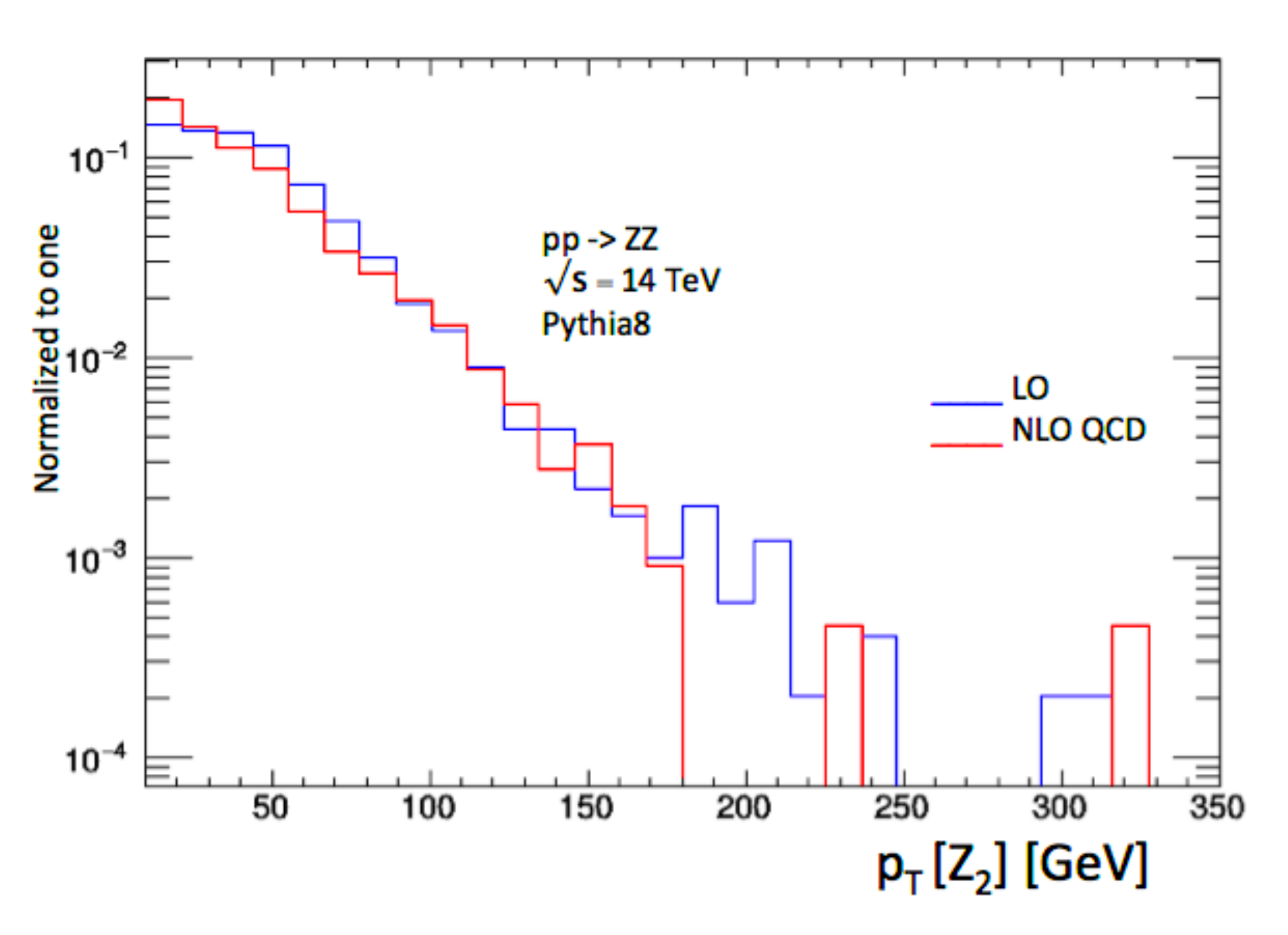}	\includegraphics[width=.30\textwidth,trim=0 0 0 0,clip]{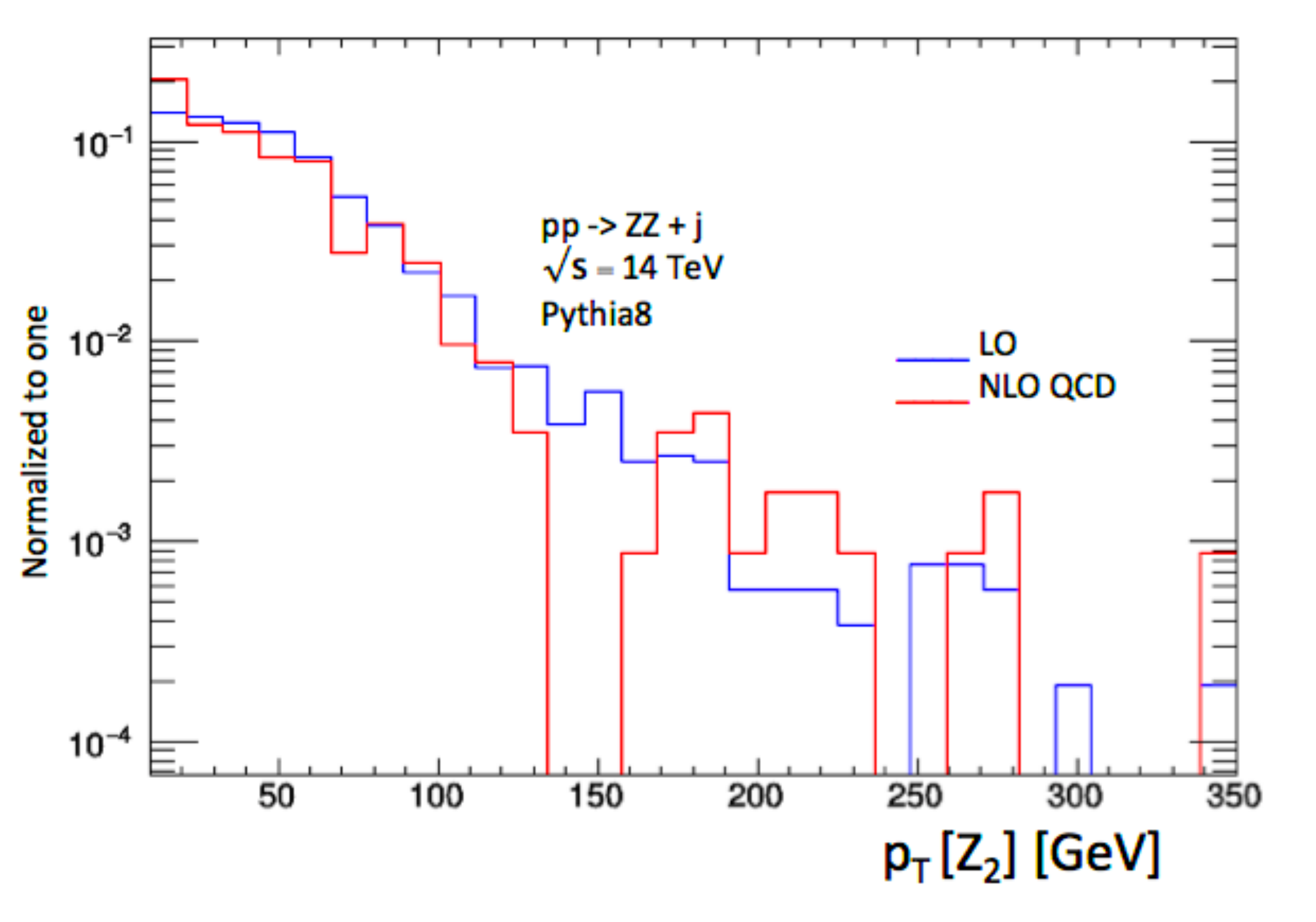} 	
	\includegraphics[width=.30\textwidth,trim=0 0 0 0,clip]{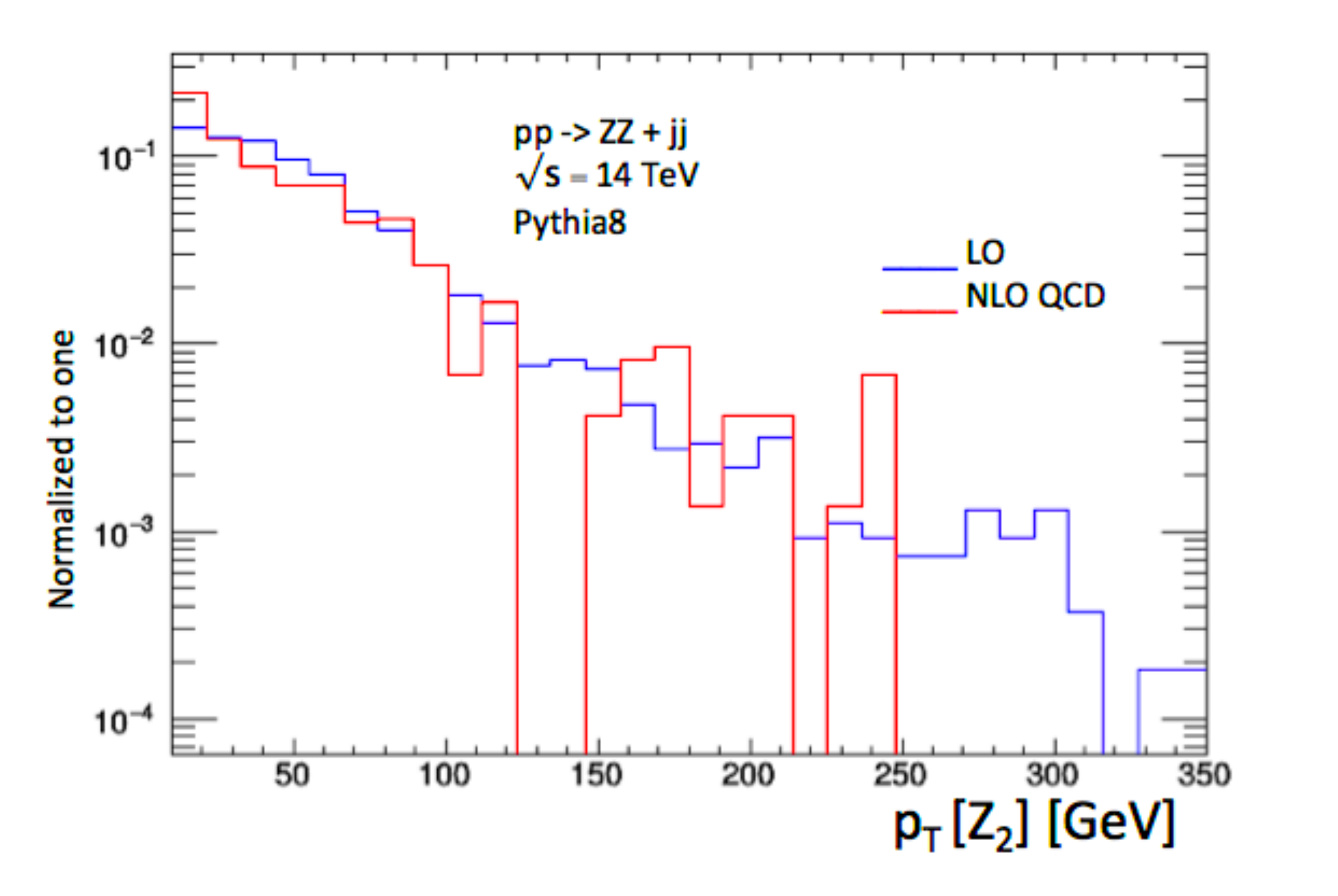} \\

	\caption{The LO and NLO transverse momentum distributions of the the subleading Z boson reconstructed after showering and hadronization with PYTHIA 8 for the processes~\ref{eq:1} at 14 TeV.
		\label{fig:9}
	}	
\end{figure}

\begin{figure}[tbp]
	\centering	
	\includegraphics[width=.30\textwidth,trim=0 0 0 0,clip]{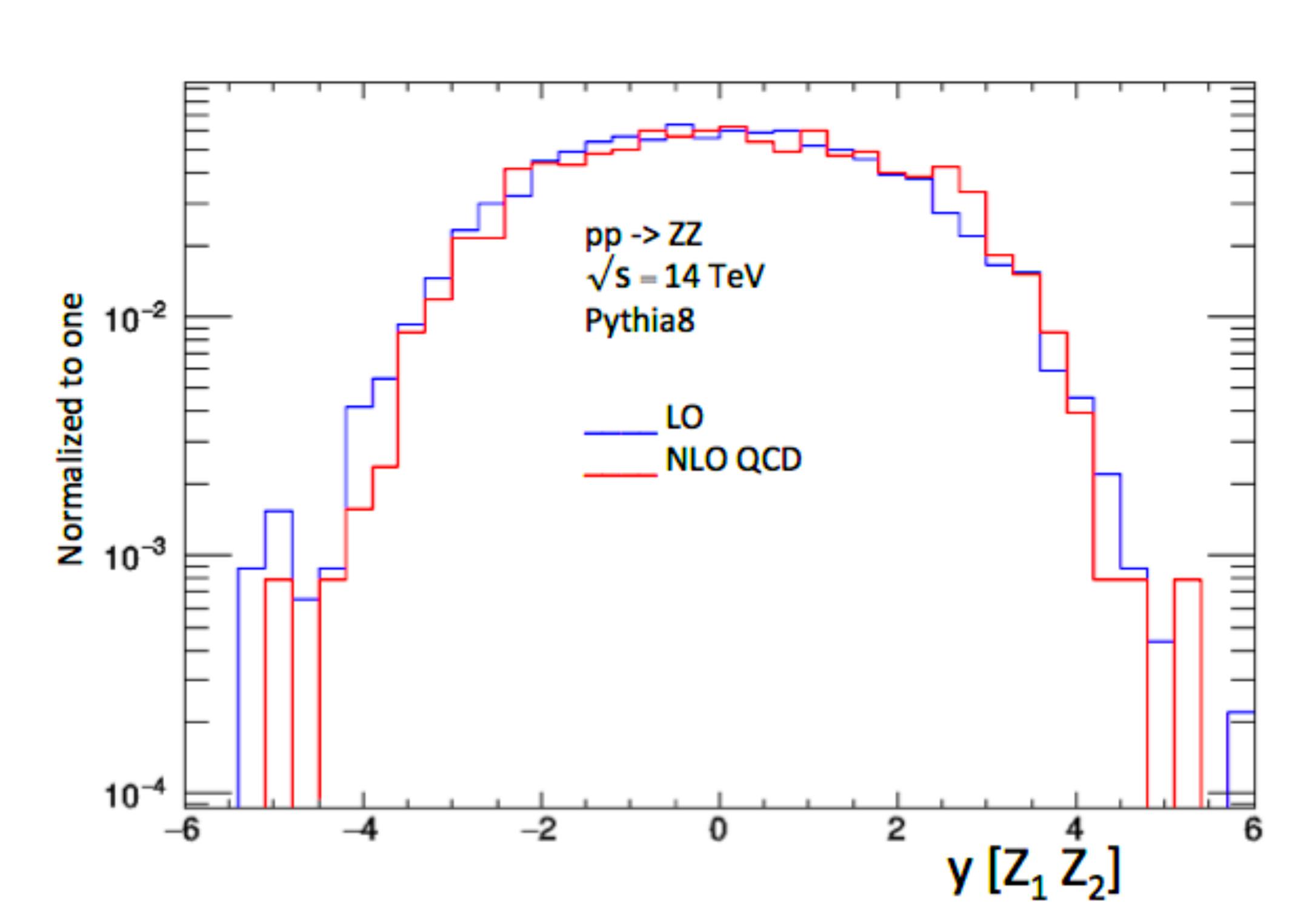}	\includegraphics[width=.30\textwidth,trim=0 0 0 0,clip]{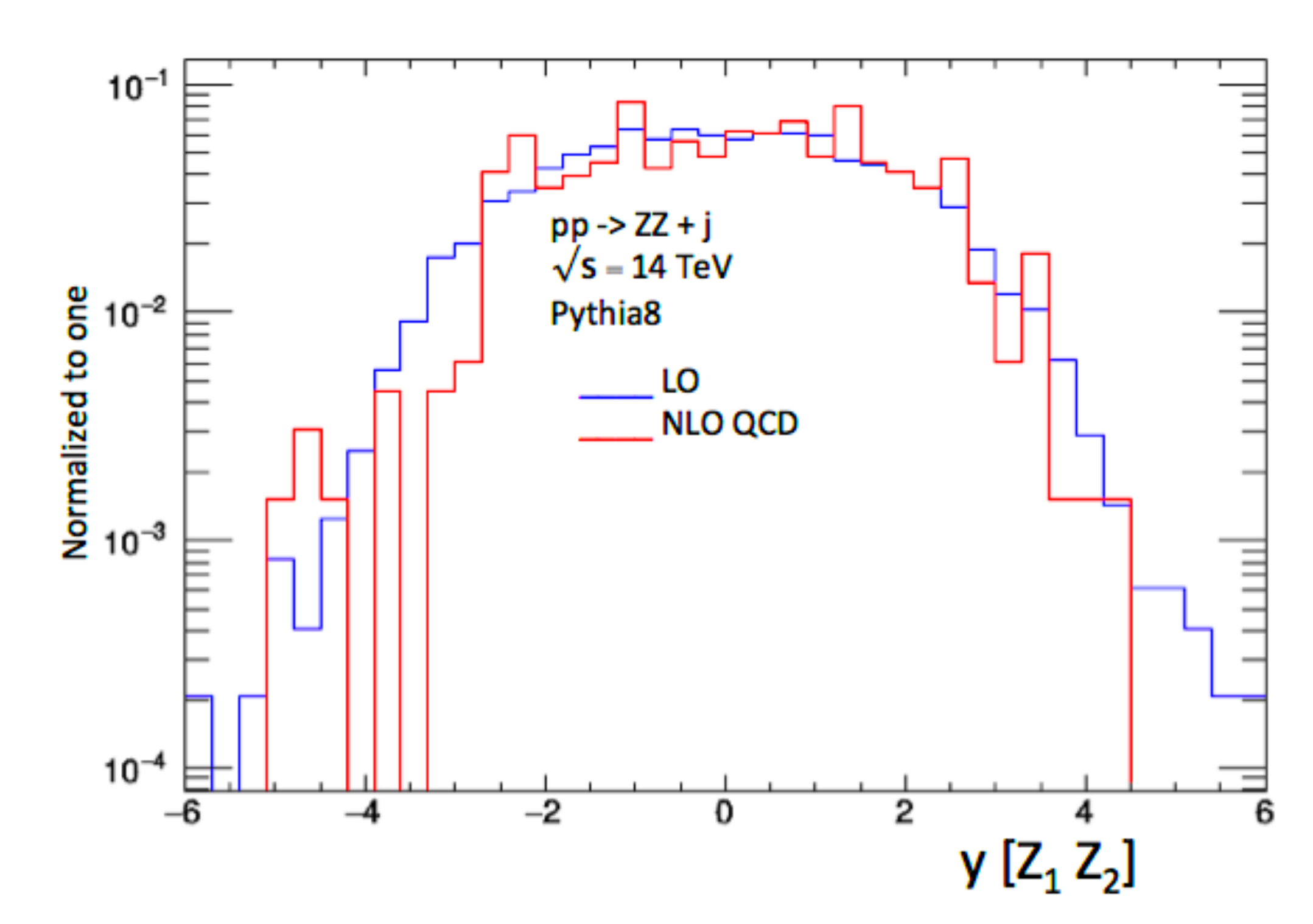} 	
	\includegraphics[width=.30\textwidth,trim=0 0 0 0,clip]{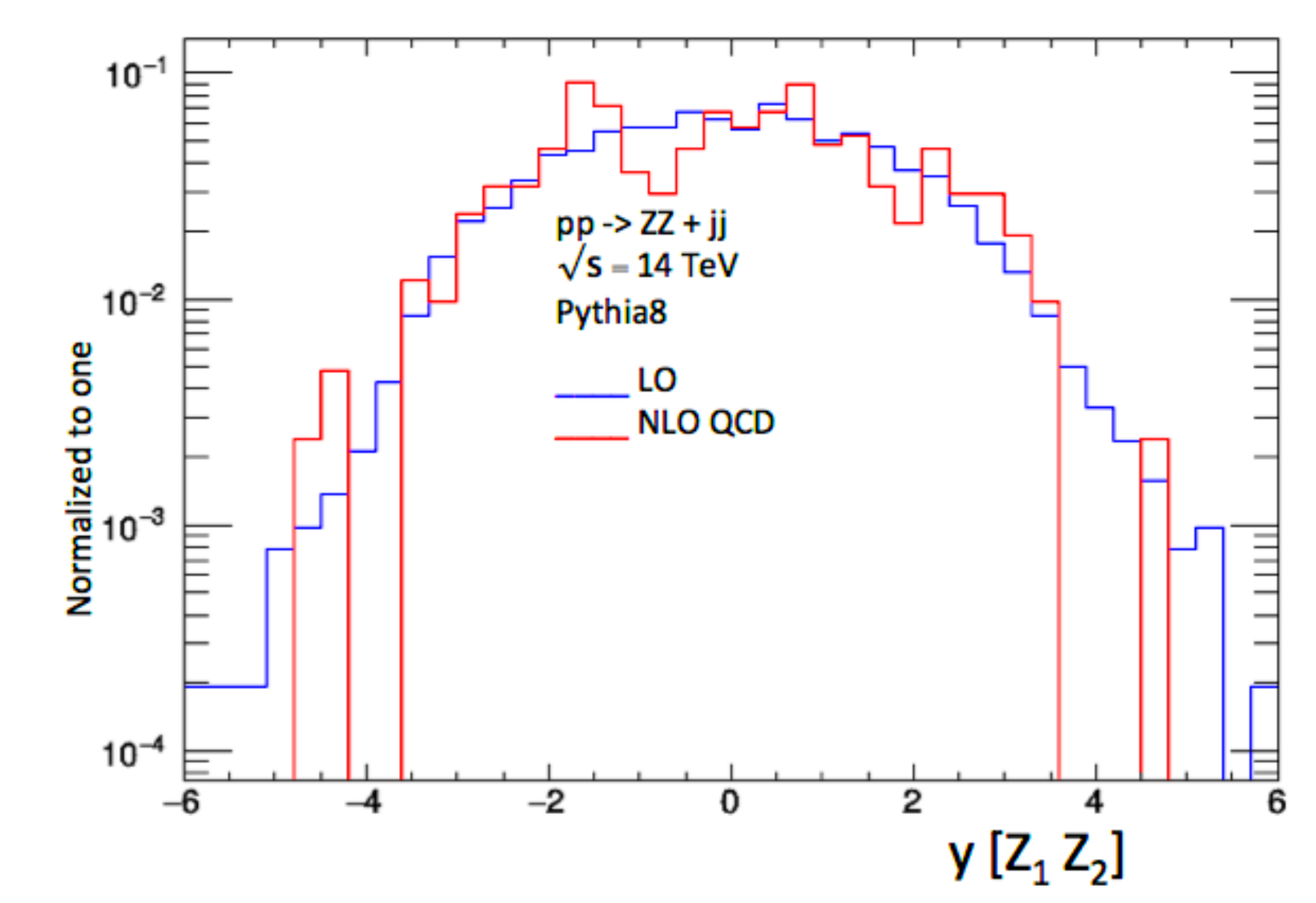} \\

	\caption{Rapidity distributions of Z-boson pair reconstructed after showering and hadronization with PYTHIA 8 for the processes~\ref{eq:1} at 14 TeV.
		\label{fig:10}
	}	
\end{figure}

\begin{figure}[tbp]
	\centering	
	\includegraphics[width=.30\textwidth,trim=0 0 0 0,clip]{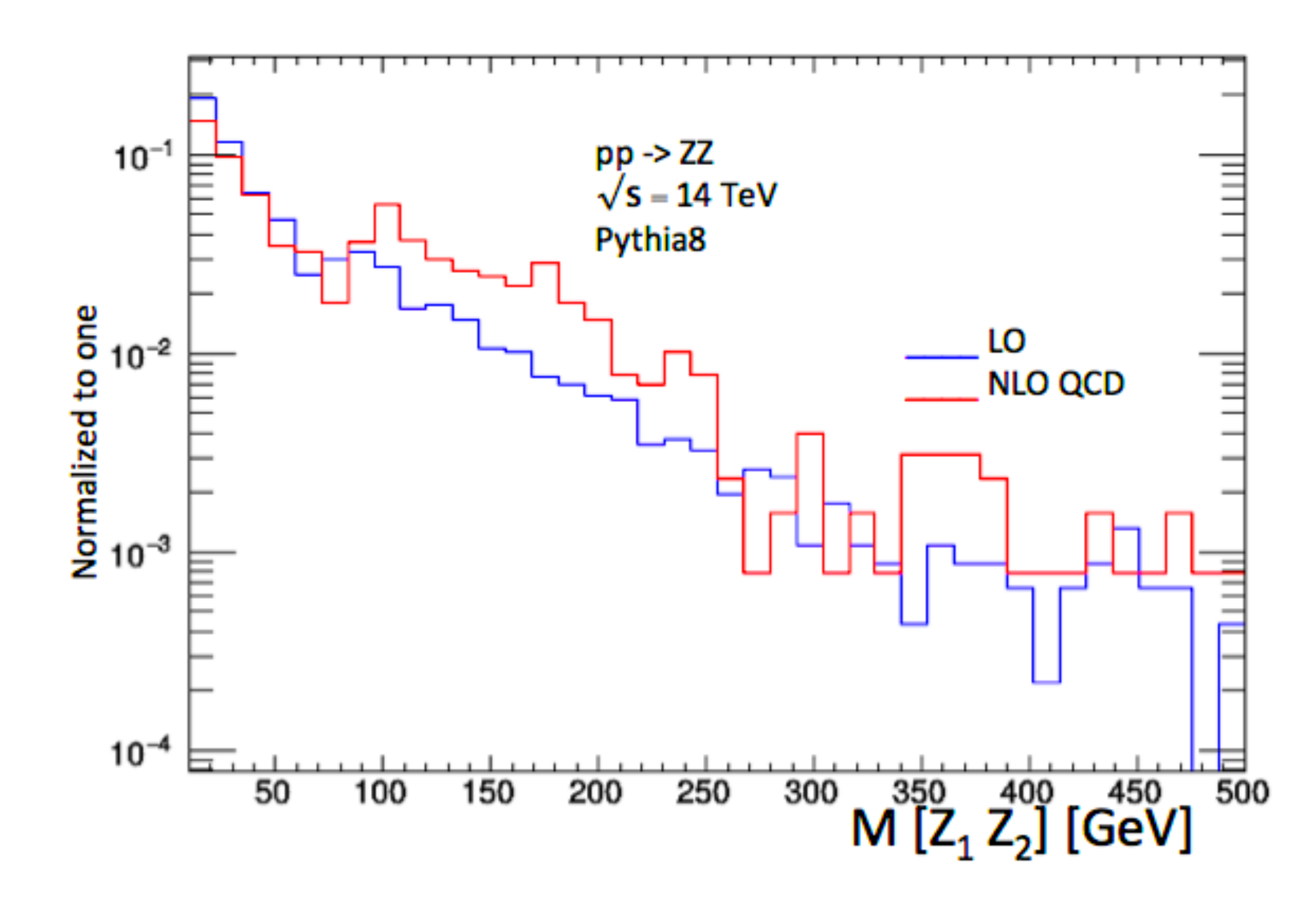}	\includegraphics[width=.30\textwidth,trim=0 0 0 0,clip]{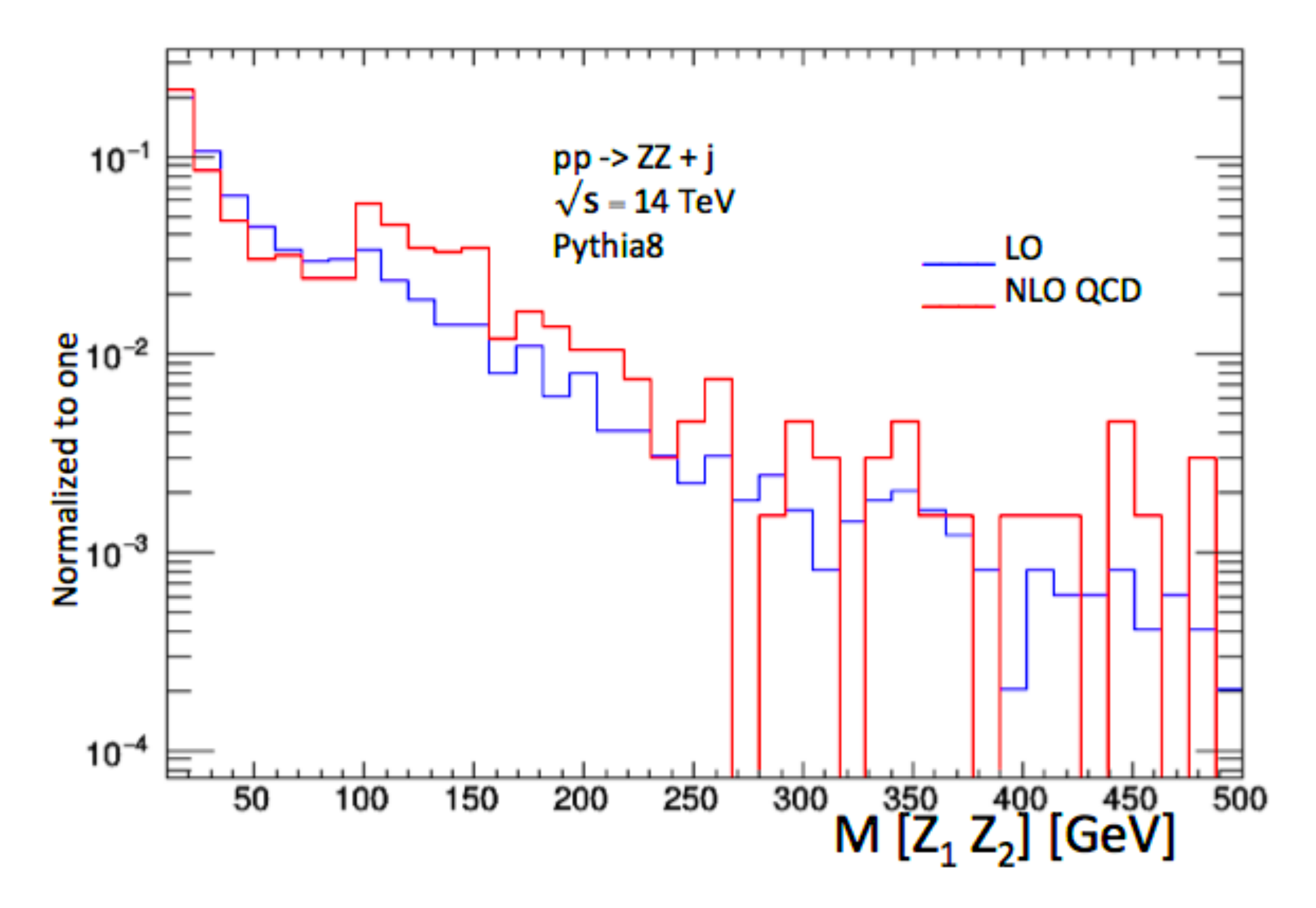} 	
	\includegraphics[width=.30\textwidth,trim=0 0 0 0,clip]{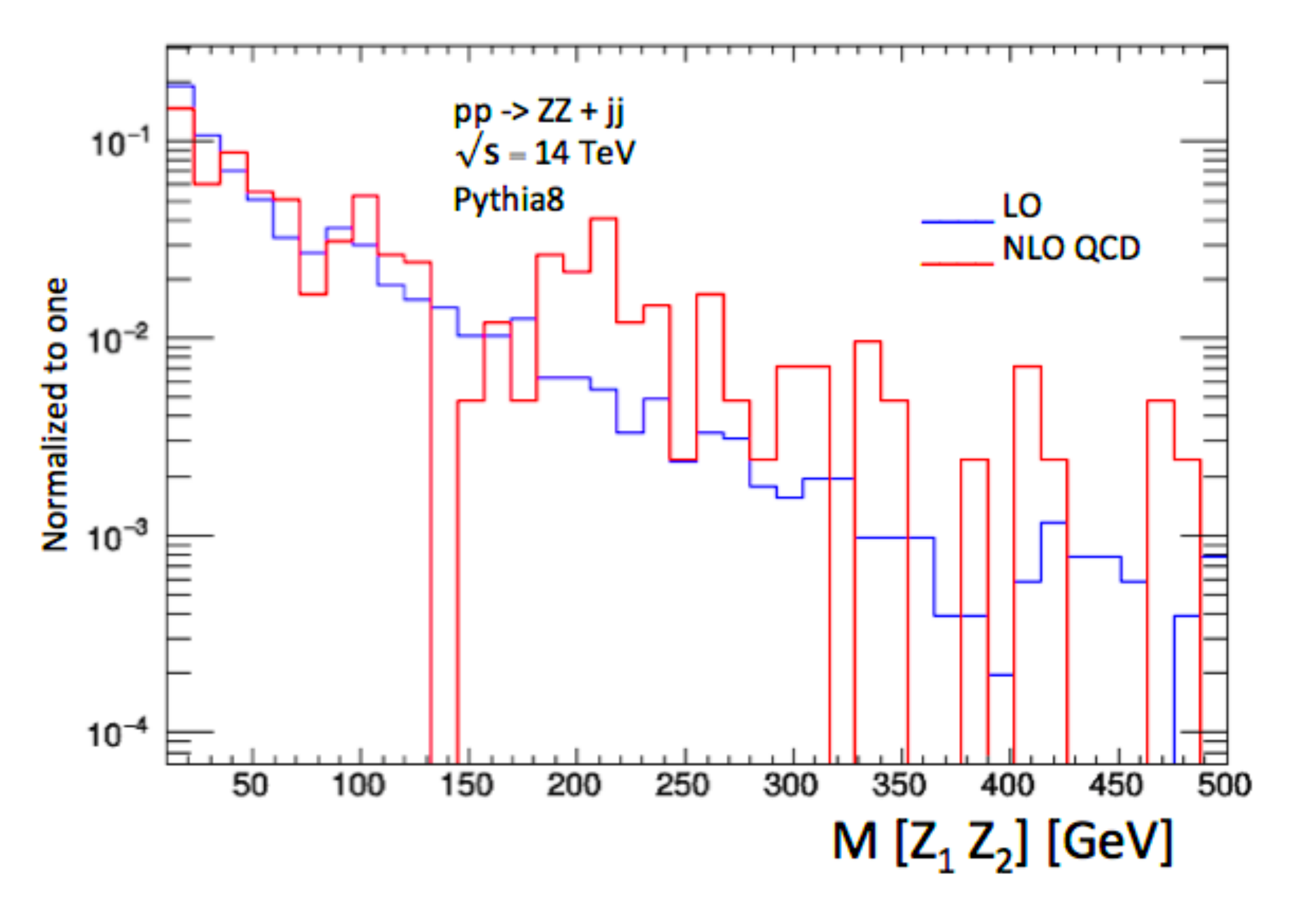} \\

	\caption{ZZ invariant mass distribution reconstructed after showering and hadronization with PYTHIA 8 for the processes~\ref{eq:1} at 14 TeV.
		\label{fig:11}
	}	
\end{figure}

We continue our discussion by presenting kinematical distributions of our processes at hadronic level, applying the showering and hadronization using Pythia8. Hence, in Figure~\ref{fig:8}, we plot the transverse momentum distribution of the leading Z boson reconstructed, we see clearly that the distribution decreases in most of the range of $p_{T} $ differently to partonic level and it vanishes rapidly in about 200 GeV, this can be explained by recoil of Z bosons with the jets in the final state. Furthermore, the jet produced by the real radiation emission influence in the absolute size of the NLO QCD corrections at low $p_{T} $.

The distribution in the $p_{T} $ of the subleading Z boson as shown in figure~\ref{fig:9}, is smaller than the distribution of the leading Z boson for all our processes, this is due to the fact that the Z bosons, the more they recede with the jets also they recede against each other, which reduces the total energy of one of the Z boson and put it as a softer Z boson.
 
Figure~\ref{fig:10} depicts the distribution in the rapidity of the two reconstructed Z bosons, we find that the parton shower has a significant improvement in $y[Z_{1} Z_{2}]$ for all processes, where we see clearly that the interval of this observable is -6 to 6, then it is wider than those of parton level, similarly the interval of the maximum for the distribution is more larger which is between -2 and 2. In addition, we note that the QCD corrections are much sizeable with parton shower but do not distort the curve shape.

Now, we going back our attention to Figure~\ref{fig:11} which shows ZZ invariant mass distribution for every processes of~\ref{eq:1}, we see a completely different behaviour from those represented at partonic level, here the distribution decreases for all values of M$_{ZZ}$, this difference can be explained by the fact that ZZ pair reconstructed from their decay into  4 leptons in the final state. From the figure we see that the QCD corrections at M$_{ZZ}$ > 90 GeV are larger, which enhance the distributions.

\section{Conclusion }
\label{sec:4}

In this paper, we have presented an implementation of ZZ, ZZ+jet and ZZ+2 jets productions at a center of mass energy of 14 TeV in proton-proton collisions at the LHC. We have generated these processes using MadGraph5$\_$aMC@NLO, at LO and NLO with QCD corrections whom give a existence to new channels by the real and the virtual emission. The NLO QCD corrections have a sizeable impact on the results of the total cross section, thus we note that they increase the LO result by an amount varying from 22$\%$ to 98$\%$. The adding of the jets to ZZ pair production in the final state decreases the total cross section, so the pp $\rightarrow $ZZ jj has a smallest value of $\sigma$. Our predictions are compared to ATLAS and CMS data in the ZZ channel at 13TeV, and we find good agreement. We have also include the one-loop-induced gluon fusion which contribute exclusively at NLO and it represents only 9$\%$ of total cross section.       
We have considered three cuts in transverse momentum of jets, thus we see  another behaviour of the total cross section: it decreases rapidly with increment of the cut, contrarily to the K factor which increases and proves that the NLO QCD corrections are significant at high transverse momentum.
 Furthermore, we have estimated the scale uncertainties by varying $\mu_{R}$ and $\mu_{F}$ simultaneously, between 0.5$M_{Z} $ < $\mu_{R}$, $\mu_{F}$ < 2$M_{Z} $ with the constraint 0.5 < $\mu_{R}$/$\mu_{F}$ < 2, we have shown that it is biggest at LO. We have also find that the  PDF uncertainties due to the  NNPDF23 LHAPDF set are less than 4$\%$.  

In the last, we have studied different kinematic distributions for all our processes at partonic level and with effect of parton shower using Pythia8.

After the showering and compared to partonic level, we observe that the shape of $p_{T}$(Z$_{1}$), $p_{T}$(Z$_{2}$) and  M$_{ZZ}$ change. Indeed,  we notice that, at the partonic level, there are two regions where the distributions have different behaviours. In the first region,  they increase then decrease progressively in the second region.  This is not true  at the hadronic level where the distributions decrease for all the values. This fact translates the parton shower effect in the processes. However this effect doesn't distort the curve of  $y[Z_{1} Z_{2}]$, but it expands its interval. 

In this study, we have presented some theoretical prediction for the on-shell ZZ, ZZ+j and ZZ+2j productions in the SM but it is interesting to use it in the context of searches for new physics  beyond the SM that we plan to do future investigation.

\section*{Acknowledgements}

This work was realized with the support of the Algerian Ministry of Higher Education and Scientic Research.



\end{document}